\documentclass[12pt, a4paper, english, russian]{extreport2}
\usepackage[cp1251]{inputenc}
\usepackage[usenames]{color}
\usepackage{colortbl}
\usepackage[T1,T2A]{fontenc}
\usepackage[russian,english]{babel}
\usepackage{amssymb,amsmath,amsfonts}
\usepackage{graphicx}
\usepackage{dcolumn}
\usepackage{bm}
\usepackage{epsfig}
\usepackage{caption}
\usepackage{cmap}

\addto\captionsrussian{\renewcommand{\figurename}{\textbf{Fig.}}}
\DeclareCaptionLabelSeparator{defffis}{ -- }
\usepackage{setspace}
\usepackage{epsfig}
\usepackage{amsmath}
\usepackage{wrapfig}
\usepackage{floatflt}
\usepackage{fancyhdr}
\usepackage{cmap}
\usepackage{indentfirst}

\usepackage{geometry}
\usepackage{fancyhdr}
\renewcommand{\thesubsection}{(\roman{subsection})}
\renewcommand{\thesubsubsection}{\roman{subsection}.}
\addto\captionsrussian{
\renewcommand\contentsname{\textit{Content}}
\renewcommand\bibname{\textit{References}}
}

\usepackage{cite}
\begin{document}
\begin{sloppy}

\setstretch{1.25}

\begin{titlepage}
\baselineskip 10pt

\begin{center}
\line(1,0){0.5}
\end{center}
\vskip 70pt
\begin{figure}[h] \centering
\epsfig{file=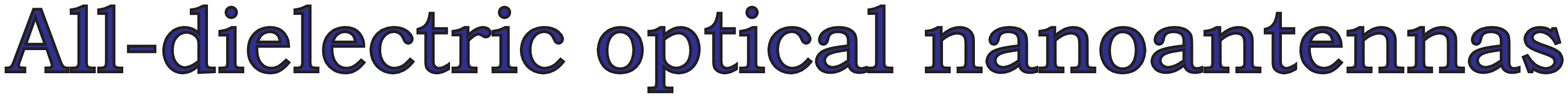,width=0.9\textwidth}
\end{figure}
\vskip 100pt
\begin{center}
Alexandr E. Krasnok, Pavel A. Belov, Andrey E. Miroshnichenko,
Arseniy I. Kuznetsov,\\ Boris S. Luk'yanchuk, Yuri S. Kivshar
\end{center}

\end{titlepage}

\begin{spacing}{1.2}
{\large \tableofcontents}
\end{spacing}

\newpage

\chapter{\textit{All-dielectric optical nanoantennas}}

\section*{Introduction}
\addcontentsline{toc}{section}{\textit{Introduction}}

Antennas are important elements of wireless information
communication technologies, along with sources of electromagnetic
radiations and their detectors. One can say that antennas are at the
heart of modern radio and microwave frequency communications
technologies. They are at the front-ends of satellites, cell-phones,
laptops and other communicating devices. In radio engineering,
antennas refer to devices converting electric and magnetic currents
into radio propagating waves and, vice versa, radio waves to
currents. Recently, the concept of antennas have been extended to
the optical frequency domain ~\cite{Novotny_09_AOP,1,
Curto_10_Science, Taminiau2008, LNov_08_Nature, 35,
Novotny_Hecht_book, 58,nanoBook}
\begin{wrapfigure}{r}{0.5\linewidth}
\centering
\includegraphics[scale=0.38]{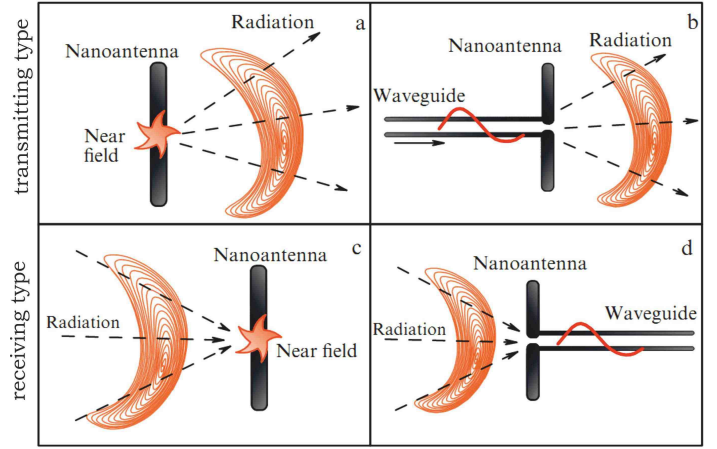}
\caption{The basic principles of nanoantenna operation (exemplified
by a nanodipole). Near field (a) or waveguide mode (b)
transformation into freely propagating optical radiation; Panels (c,
d) illustrate a reception regime. The configuration of feeding via a
plasmonic waveguide is of great importance for practical
applications of nanoantennas, especially for the development of
wireless communication systems at the nanometer level, i.e., for
future photonic chips.~\cite{58}} \label{pic1}
\end{wrapfigure}

as a result of the development of a
new branch of physics emerged known as nanooptics, which studies the
transmission and reception of optical signals by submicron and even
nanometer-sized objects.

For nanooptics it is important to efficiently detect and direct the
transmitting signals for optical information between nanoelements.
The sources and detectors of radiation in nanooptics are
nanoelements themselves, their clusters, and even individual
molecules (atoms, ions). Nanoobjects functioning as antennas must
exhibit high radiation efficiency and directivity.

Nanoantennas, similar to the radiofrequency antennas, are usually
divided into two types, transmitting and receiving
(see Fig. \ref{pic1}). Figure \ref{pic1}a schematically shows the
interaction between a nanoantenna and the near field of an quantum
emitter. In this case, the nanoantenna transforms the near field
into freely propagating optical radiation, i.e. it is a
transmitting nanoantenna. Figure \ref{pic1}c illustrates the
operation of a receiving nanoantenna that converts incident
radiation into a strongly confined near field.

The energy is usually delivered to a microwave antenna through a
waveguide. Such an antenna converts waveguide modes to freely
propagating radiation. In the case of optical antennas with their
sufficiently small optical size, the waveguide mode must have the
subwavelength cross section attainable by using so-called plasmonic
waveguides. This type of nanoantenna feeding is depicted
schematically in Fig. \ref{pic1}b. According to the reciprocity principle,
such a nanoantenna is also capable of transforming
incident radiation to plasmonic waveguide modes (see Fig. \ref{pic1}d).

\begin{wrapfigure}{r}{0.5\linewidth}
\centering
\includegraphics[scale=0.45]{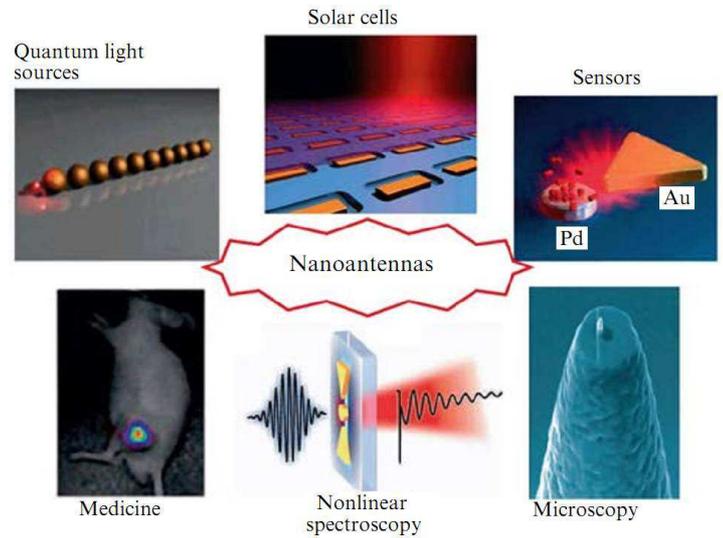}
\caption{Plethora of nanoantennas application in modern
science.~\cite{58}} \label{pic2}
\end{wrapfigure}

Thus, the transmitting antenna converts a strongly confined field in
the optical frequency range created by a certain (weakly emitting or
almost non emitting) source into optical radiation (see Fig.
\ref{pic1}a,b). Conversely, the receiving nanoantenna is a device
efficiently converting incident light (optical frequency radiation)
into a strongly confined field (see Fig. \ref{pic1}c,d), where an electromagnetic field
is concentrated in a small region compared to the wavelength of light.
Such fields are characterized by a spatial spectrum consisting mostly of evanescent waves.
The confinement region may be of subwavelength dimension, leading to a strongly
confined near field. The energy of this field contains contributions from stored and non radiated energy. However, an important particular case of nanoantennas is a device converting
optical radiation into waveguide modes, and vice versa, as shown in
Fig. \ref{pic1}c,d. In this case, the subwavelength dimension is
characterized by the transverse cross section of the strongly
confined field region. The longitudinal size of this region (along
the waveguide axis) may be optically large, and the electromagnetic
energy of the strongly confined field is referred to as expanding.
The feeding configuration with a plasmonic waveguide is of great
importance for practical applications of nanoantennas, especially
for the development of wireless communication systems at the
nanometer level, i.e., for future fully optical integrated circuits.

\begin{figure}[!b]
\center{\includegraphics[width=1\linewidth]{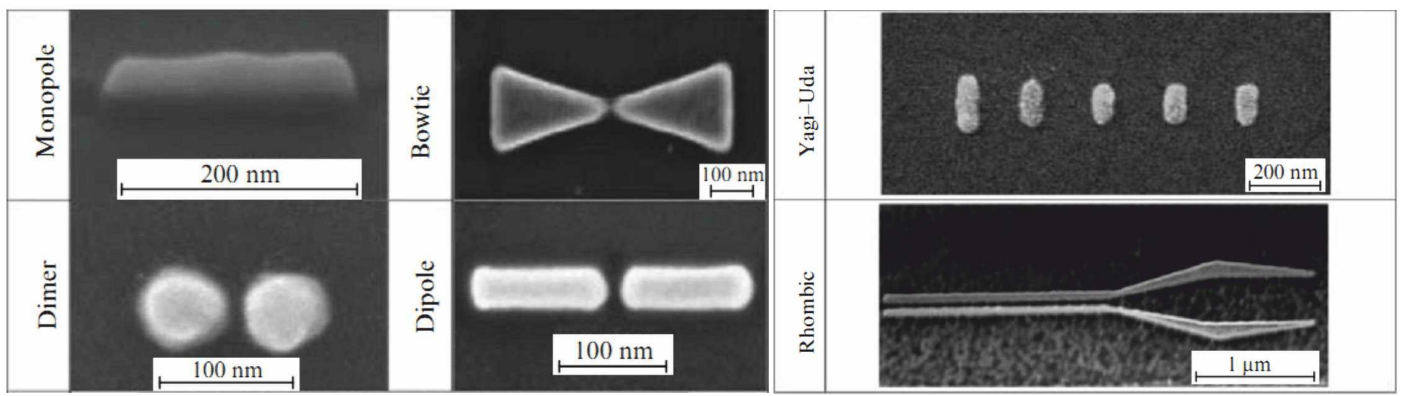}} \caption{Main
types of plasmonic nanoantennas.~\cite{58}} \label{pic3}
\end{figure}

Nanoantennas are the most promising area of research in the modern
nanooptics due to their ability to bridge the size and
impedance mismatch between nanoemitters and free space radiation, as
well as manipulate light on the scale smaller than the wavelength of
light. Bearing in mind the great variety of sources and detectors of
strongly confined optical fields (groups of atoms and molecules,
luminescent and fluorescent cells, e.g., viruses and bacteria,
sometimes individual molecules, quantum dots, and quantum wires), it
is safe to say that the areas of practical applications of
nanoantennas in the near future will be commensurate with that of
their classical analogs.

At present, nanoantennas are used in near-field microscopy and
high-resolution biomedical sensors; their application for
hyperthermal therapy of skin neoplasms is a matter of the
foreseeable future. There are some other potential applications of
nanoantennas (see Fig. \ref{pic2}) that we believe to be equally
promising, including solar cells ~\cite{2}, molecular and biomedical
sensors ~\cite{3}, optical communication ~\cite{4}, and optical
tweezers ~\cite{5}. The variety of applications allows us to argue
that the concept of nanoantennas presents a unique example of the
penetration of new physics into various spheres of human activity.

Thus far, optical antennas have primarily been constructed from
metallic materials, which support plasmonic resonances. The main
types of plasmonic nanoantennas which have been realized
experimentally are presented in Fig. \ref{pic3}. Different types of
plasmonic nanoantennas are designed to perform various tasks. For
example dipole nanoantennas~\cite{58, nanoBook,dip_v_OE_12,
Khlebtsov,Gonzalez05} demonstrate high coefficient of electric field
localization, while bowtie nanoantennas~\cite{58, nanoBook,
NL_bowtie_dip_11, Guo_08, Hatab_10, Moerner_bowtie_09, Toussaint_11,
5, Sederberg_11, Odom_12, Schuck_09} are broadband; Yagi-Uda type
nanoantennas exhibit high directivity which is very useful for
optical wireless communications on an optical
chip~\cite{Curto_10_Science,58,nanoBook, 4, Alu_dip_PRL_08,
ALU_NP_08, Koenderink_11_NLetters, Dorfmuller_11_NLetters,
Kadoya_NJF_07, Lerosey_NP_10, Engheta_09_PRB, Lobanov12, 20, 21,
Hulst_08_OSA}. However, despite of a number of advantages of
plasmonic nanoantennas associated with their small size and strong
localization of the electric field, such nanoantennas have large
dissipative losses resulting in low radiation efficiency.

To overcome such limitations, we propose a new type of nanoantennas
based on dielectric nanoparticles with a high index dielectric
constant~\cite{KrasnokNanoscale, KrasnokOE, 44, 43, nanoBook, 58, Krasnok_12AIP, SPIEKrasnok, Noskov12Arxiv, KrasnokOligomerAPL, Krasnok_11}, for example Huygens optical elements and
Yagi-Uda nanoantennas [see par.(\ref{HuyYgi})]. Such all-dielectric
optical nanoantennas will have low dissipative losses with enhanced
magnetic response in the visible. The concept of optical magnetism
based on dielectric nanoparticles is presented in the next section.
The key for such novel functionalities of high index dielectric
nanophotonic elements is the ability of subwavelength dielectric
nanoparticles to support simultaneously both electric and magnetic
resonances, which can be controlled independently. This type of
nanoantennas has several unique features such as low optical losses
at the nanoscale and superdirectivity. The concept of all-dielectric
nanoantennas has been developed in our original papers~\cite{58,
7,8, Krasnok_12AIP, SPIEKrasnok, Noskov12Arxiv, nanoBook} and also
summarized below.

Furthermore all-dielectric nanoantennas allow us achieve the
superdirectivity effect. Superdirectivity as a physical concept can
be found in textbooks on antennas, however all so far proposed superdirective antennas are not reliably reproducible. More specifically, all previous attempts to achieve superdirectivity of
antennas were based on discrete arrays of radiating dipoles with a
rather cumbersome distribution of radiating currents over the array.
This approach resulted in intrinsic drawbacks of known superdirective arrays - ultra-narrowfrequency range, high dissipation, and extreme sensitivity to any disturbance, etc. As a
result, no single superdirective antenna was demonstrated up to now.
In the context of nanoantennas, which originated from radio frequency antennas a few years ago, superdirectivity has never been discussed. However, superdirectivity would be a very desirable
feature in nanophotonics with numerous useful applications. Here we
describe [see par.(\ref{SupDirect})] the superdirectivity effect in
a very simple, elegant, and practical way for a nanoparticle with a
notch. This approach is able to shape higher-harmonics of the radiation field
in such a way that not only superdirectivity of this nanoantennas
becomes possible but also a strong subwavelength sensitivity of the radiation pattern to the location of the emitter can be easily realized.

\section{Optical magnetism based on dielectric nanoparticles}

\begin{figure}[!t] \centering
\includegraphics[width=0.7\textwidth]{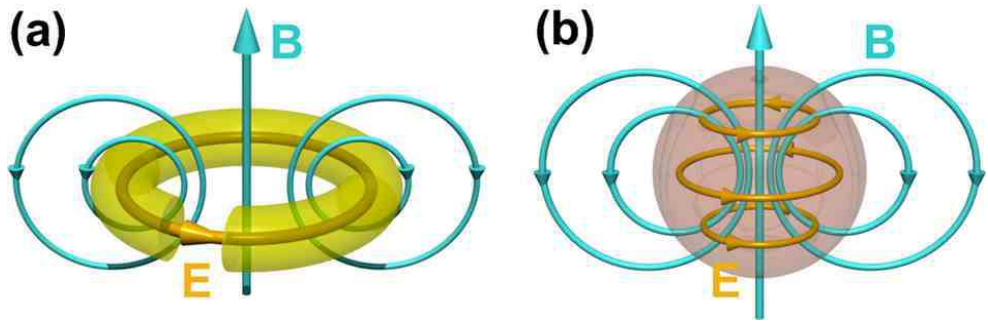}
\caption{Schematic representation of electric and magnetic field
distribution inside a metallic split-ring resonator (a) and a
high-refractive index dielectric nanoparticle (b) at magnetic
resonance wavelength.~\cite{10}} \label{SRRandParticle}
\end{figure}

It is well known that a pair of oscillating electric charges of
opposite signs, know as an oscillating electric dipole, produces
electromagnetic radiation at the oscillations frequency ~\cite{Landau}.
Although, distinct “magnetic charges”, or monopoles, have not been observed so far, magnetic dipoles are very common sources of magnetic field in nature. The field of the
magnetic dipole is usually calculated as the limit of a current loop
shrinking to a point. Its profile is equivalent to the one of an electric dipole considering that the electric and magnetic fields are exchanged. The most common example
of a magnetic dipole radiation is an electromagnetic wave produced
by an excited metal split-ring resonator (SRR), which is a basic
constituting element of metamaterials (see Fig. \ref{SRRandParticle}a)~\cite{Shalaev, Zheludev, Soukoulis,
Boltasseva, Soukoulis1, Pendry, Smith, Shelby, Smith1}. The real
currents excited by external electromagnetic radiation and running
inside the SRR produce a transverse oscillating up and down magnetic
field in the center of the ring, which simulates an oscillating
magnetic dipole. The major interest of these artificial systems is
due to their ability to response to a magnetic component of incoming
radiation and thus to have a non-unity or even negative magnetic
permeability ($\mu$) at optical frequencies, which does not exist in
nature. This provides possibilities to design unusual material
properties such as negative refraction ~\cite{Shalaev, Zheludev,
Soukoulis, Boltasseva, Soukoulis1, Pendry, Smith, Shelby, Smith1},
cloaking ~\cite{Leonhardt, Pendry06}, or superlensing~\cite{Pendry00}.
The SRR concept works very well for gigahertz~\cite{Smith, Shelby,
Smith1}, terahertz~\cite{Yen06, Padilla06} and even near-infrared
(few hundreds THz)~\cite{Linden04, Enkrich05, Liu08} frequencies.
However, for shorter wavelengths and in particular for visible
spectral range this concept fails due to increasing losses and
technological difficulties to fabricate smaller and smaller
constituting split-ring elements~\cite{Enkrich05, Soukoulis07}.
Several other designs based on metal nanostructures have been
proposed to shift the magnetic resonance wavelength to the visible
spectral range~\cite{Shalaev, Zheludev}. However, all of them are
suffering from losses inherent to metals at visible frequencies.

An alternative approach to achieve strong magnetic response with low
losses is to use nanoparticles made of high-refractive index
dielectric materials~\cite{Soukoulis1,Zhao09}. As it follows from the
exact Mie solution of light scattering by a spherical particle,
there is a particular parameter range where strong magnetic dipole
resonance can be achieved. Remarkably, for the refractive indices
above a certain value there is a well-established hierarchy of
magnetic and electric resonances. In contrast to plasmonic particles
the first resonance of dielectric nanoparticles is a magnetic dipole
resonance, and takes place when the wavelength of light inside the
particle equals to the diameter $\lambda/n_{s}\simeq2 R_{s}$, where $\lambda$ is a wavelength in a free space, $R_{s}$ and $n_{s}$ are the radius and refractive index of spherical particle.
Under this condition the polarization of the electric field is
anti-parallel at opposite boundaries of the sphere, which gives rise
to strong coupling to circulation displacement currents while
magnetic field oscillates up and down in the middle (see Fig.
\ref{SRRandParticle}b).

\begin{figure}[!b] \centering
\includegraphics[width=0.7\textwidth]{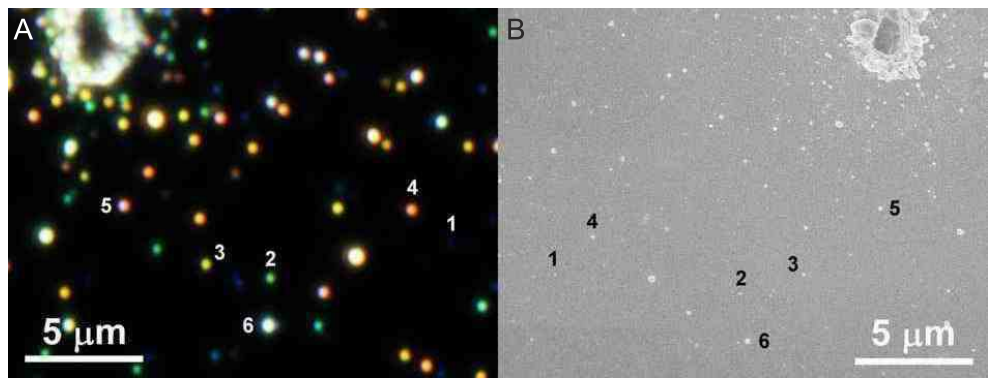}
\caption{Dark-field microscope (a) and top-view scanning electron
microscope (SEM) (b) images of the same area on a silicon wafer
ablated by a femtosecond laser. Microscope image is inverted in
horizontal direction relative to that of the SEM. Selected
nanoparticles are marked by corresponding numbers 1 to 6 in both
figures.~\cite{10}} \label{MagLight}
\end{figure}

Below in this section we present the experimental results
demonstrating ~\cite{10} that spherical silicon nanoparticles with
sizes in the range from 100 nm to 200 nm have strong magnetic dipole
response in the visible spectral range. The scattered “magnetic”
light by these nanoparticles is so strong that it can be easily seen
under a dark-field optical microscope. The wavelength of this
magnetic resonance can be tuned throughout the whole visible
spectral range from violet to red by just changing the nanoparticle
size.

In article ~\cite{10} we have chosen silicon (Si) as a material which
has high refractive index in the visible spectral range (above 3.8
at 633 nm) on one side and still almost no dissipation losses on
the other. Silicon nanorods have attracted considerable attention
during the last few years due to their ability to change their
visible color with the size~\cite{Cao10}. This effect appears due to
excitation of particular modes inside the cylindrical silicon
nanoresonators. Moreover, recent theoretical work predicted that
spherical silicon nanoparticles with sizes of a few/several hundred
nanometers should have both strong magnetic and electric dipole
resonances in the visible and near-IR spectral range ~\cite{9,53}. To
fabricate the silicon nanoparticles we have used the laser ablation
technique, which is an efficient method to produce nanoparticles of
various materials and sizes~\cite{49}. Nanoparticles produced by the
ablation method can be localized on a substrate and measured
separately from each other using single nanoparticle spectroscopy.

\begin{figure}[!b] \centering
\includegraphics[width=0.8\textwidth]{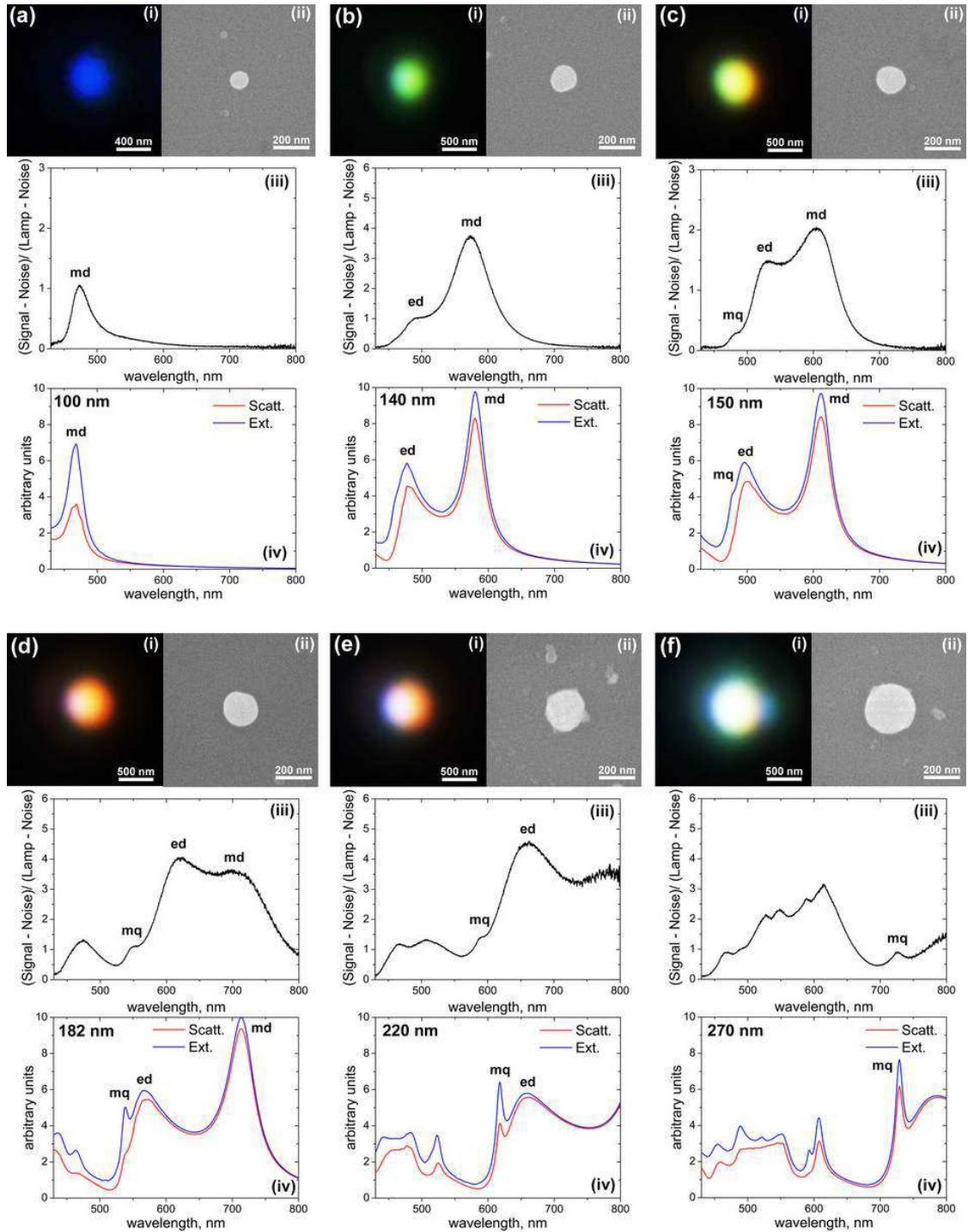}
\caption{Close-view dark-field microscope (i) and SEM (ii) images of
the single nanoparticles selected in Fig. \ref{MagLight}. Figures
(a) to (f) correspond to nanoparticles 1 to 6 from Fig.
\ref{MagLight} respectively. (iii) Experimental dark-field
scattering spectra of the nanoparticles. (iv) Theoretical scattering
and extinction spectra calculated by Mie theory for spherical
silicon nanoparticles of different sizes in free space.
Corresponding nanoparticle sizes are defined from the SEM images
(ii) and noted in each figure.~\cite{10}} \label{scatMiExper}
\end{figure}

Dark-field microscopic image of a silicon sample ablated by a
focused femtosecond laser beam is shown in Fig. \ref{MagLight}a. It
shines by all the colours of the rainbow from violet to red. To
clarify the origin of this strong scattering we selected some
nanoobjects shining with different colours on the sample (see Fig.
\ref{MagLight}a) and measured their scattering spectra by single
nanoparticle dark-field spectroscopy. Then, the same sample area was
characterized by scanning electron microscopy and the selected
nanoobjects providing different colours have been identified (see Fig.
\ref{MagLight}b, the dark-field microscope image is inverted in
horizontal direction relative to that of the SEM). The results of
this comparative analysis of the same nanoobjects by dark-field
optical microscopy, dark-field scattering spectroscopy, and scanning
electron microscopy are presented in Fig. \ref{scatMiExper}. As it
can be seen from the SEM images the observed colours are provided by
silicon nanoparticles of almost perfect spherical shape and varied
sizes. This makes it possible to analyze scattering properties of
these nanoparticles in the frames of Mie theory~\cite{52} and
identify the nature of optical resonances observed in our spectral
measurements. The bottom panels (iv) in Fig. \ref{scatMiExper}
represent a total extinction cross-section calculated using Mie
theory~\cite{52} for silicon nanoparticles of different sizes (the
calculations were done in free space). In these calculations, the
size of the nanoparticles in each figure was chosen to be similar to
the size defined from each corresponding SEM image (ii). It can be
seen that there is a clear correlation between the experimental
(iii) and theoretical spectra (iv) both in the number and position
of the observed resonances. This makes it obvious that Mie theory
describes more or less accurately our experimental results.

One of the main advantages of the analytical Mie solution compared
to other computational methods is its ability to split the observed
spectra into separate contributions of different multipole modes and
have a clear picture of the field distribution inside the particle
at each resonance maximum. This analysis was done for each particle
size in Fig. \ref{scatMiExper} and corresponding multipole
contributions were identified (see notations in the experimental and
theoretical spectra). According to this analysis the first strongest
resonance of these nanoparticles appearing in the longer wavelength
part of the spectrum corresponds to magnetic dipole response (md).
Electric field inside the particle at this resonance wavelength has
a ring shape while magnetic field oscillates in the particle center.
Magnetic dipole resonance is the only peak observed for the smallest
nanoparticles (see Fig. \ref{scatMiExper}a). At increased nanoparticle
size (see Fig. \ref{scatMiExper}b,c) electric dipole (ed) resonance also
appears at the blue part of the spectra, while magnetic dipole
shifts to the red. For relatively small nanoparticles, the observed
colour is mostly defined by the strongest resonance peak and changes
from blue to green, yellow, and red when magnetic resonance
wavelength shifts from 480 nm to 700 nm (see Fig. \ref{scatMiExper}a–d).
So, we can conclude that the beautiful colours observed in the dark
field microscope (see Fig. \ref{MagLight}a) correspond to magnetic
dipole scattering of the silicon nanoparticles, “magnetic light”.
Further increase of the nanoparticle size leads to the shift of
magnetic and electric dipole resonances further to the red and
infra-red frequencies, while higher multipole modes such as magnetic
and electric quadrupoles appear in the blue part of the spectra
(see Fig. \ref{scatMiExper}d–f).

Some differences between experimental and theoretical spectra
observed in Fig. \ref{scatMiExper} can be attributed to the presence
of silicon substrate, which is not taken into account in our simple
Mie theory solution. We should also mention that very similar
results have been published almost simultaneously by a different
group of authors ~\cite{11} who demonstrated magnetic and electric
dipole resonances of silicon particles in red and near-IR spectral
range.

Recently we have also experimentally demonstrated for the first time
directional light scattering by spherical silicon nanoparticles in
the visible spectral range ~\cite{17}. These unique scattering
properties arise due to simultaneous excitation and mutual
interference of magnetic and electric dipole resonances inside a
single nanosphere. This phenomenon is similar to a known since long
time Kerker-type scattering predicted in~\cite{Kerker:JOSA:1983} for
hypothetical magneto-dielectric nanoparticles but never observed
experimentally. Directivity of the far-field radiation pattern can
be controlled by changing light wavelength and the nanoparticle
size. Forward-to-backward scattering ratio above 6 was
experimentally obtained at visible wavelengths. Similar directional
light scattering by spherical ceramic particles in GHz
~\cite{Geffrin} and GaAs nanodisks in the visible
~\cite{Person:NL:2013} has also been published almost simultaneously
by different groups of authors. These unique optical properties of
high-refractive index dielectric nanostructures constitute the
background for our approach to all-dielectric nanoantennas, which
will be discussed in detail below.

\section{Huygens optical elements and Yagi—Uda nanoantennas based on dielectric
nanoparticles}\label{HuyYgi}

Recently, it was suggested~\cite{58, 7, 8, Krasnok_12AIP,
SPIEKrasnok, Noskov12Arxiv, nanoBook} a novel type of optical
nanoantennas made of all-dielectric elements. Moreover, we argue
that, since the source of electromagnetic radiation is applied
externally, dielectric nanoantennas can be considered as the best
alternative to their metallic counterparts. First, dielectric
materials exhibit low loss at the optical frequencies. Second, as
was suggested earlier, nanoparticles made of high-permittivity
dielectrics may support both electric and magnetic resonant modes.
This feature may greatly expand the applicability of optical
nanoantennas for, e.g. for detection of magnetic dipole transitions of
molecules~\cite{Schmidt12}. In our study we concentrate on
nanoparticles made of silicon. The real part of the permittivity of
the silicon in the visible spectral range is about $16$~
\cite{Palik}, while the imaginary part is up to two orders of
magnitude smaller than that of nobel metals (silver and gold).

\subsection{General concept}

\begin{figure}[!t] \centering
\includegraphics[width=0.65\textwidth]{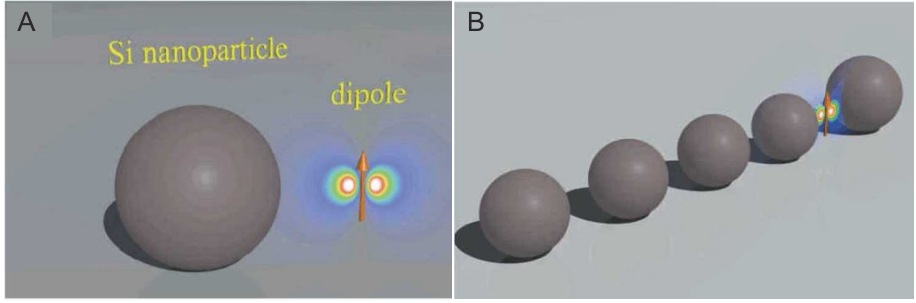}
\caption{({\bf A}) Huygens element consisting of a single silicon
nanoparticle and point-like dipole source separated by a distance
$G_{ds}=90$ nm (between dipole and sphere surface). The radius of
the silicon nanoparticle is $R_{s}=70$ nm. ({\bf B}) Dielectric
optical Yagi-Uda nanoantenna, consisting of the reflector of the
radius $R_r = 75$ nm, and smaller director of the radii $R_d = 70$
nm. The dipole source is placed equally from the reflector and the
first director surfaces at the distance G. The separation between
surfaces of the neighbouring directors is also equal to G.~\cite{7}}
\label{fig1}
\end{figure}

The mentioned above properties of dielectric nanoparticles allow us to
realize optical Huygens source ~\cite{29} consisting of a point-like
electric dipole operating at the magnetic resonance of a dielectric
nanosphere (see Fig.\ref{fig1}A). Such a structure exhibits high
directivity with vanishing backward scattering and polarization
independence, being attractive for efficient and compact designs of
optical nanoantennas.

We start our analysis by considering a radiation pattern of two
ideal coupled electric and magnetic dipoles. A single point-like
dipole source generates the electric far-field of the following
form

\begin{equation}
\mathbf{E}_p =
\frac{k^2}{4\pi\epsilon_0r}\exp(ikr)\left[\mathbf{p}-\mathbf{n}(\mathbf{n}\cdot\mathbf{p})\right],
\end{equation}
where $\mathbf{p}$ is the electric dipole, $k = \omega/c$ is
the wavenumber, $\mathbf{n}$ is the scattered direction, and
$r$ is the distance from the dipole source. The radiation pattern
$\sigma=\lim\limits_{r\rightarrow\infty}4\pi r^2|E_p|^2$ in the
plane of the dipole $\mathbf{n}\times\mathbf{p}=0$ is
proportional to the standard figure-eight profile,
$\sigma_{||}\propto|\cos\alpha|^2$, where $\alpha$ is the scattered
angle. In the plane orthogonal to the dipole
($\mathbf{n}\cdot\mathbf{p}=0$) the radiation pattern
remains constant and angle independent, $\sigma_{\bot}\propto
const$. Thus, the total radiation pattern of a single dipole emitter
is a torus which radiates equally in the opposite directions. If we
now place, in addition to the electric dipole, an orthogonal
magnetic dipole located at the same point, the situation changes
dramatically. The magnetic dipole $\mathbf{m}$ generates the
electric far-field of the form

\begin{equation}
\mathbf{E}_m =
-\sqrt{\frac{\mu_0}{\epsilon_0}}\frac{k^2}{4r\pi}\exp(ikr)\left(\mathbf{n}\times\mathbf{m}\right).
\end{equation}
Thus, the total electric field is a sum of {\em two contributions}
from both electric and magnetic dipoles $\mathbf{E}_{\rm total} =
\mathbf{E}_p+\mathbf{E}_m$. By assuming that the magnetic dipole is
related to the electric dipole via the relation
$|\mathbf{m}|=|\mathbf{p}|/(\mu_0\epsilon_0)^{1/2}$,
which corresponds to an infinitesimally small wavefront of a plane
wave often called a Huygens source~\cite{29}, the radiation pattern
becomes $\sigma^H\propto|1+\cos\alpha|^2$. This radiation pattern is
quite different compared to that of a single electric dipole. It is
{\em highly asymmetric} with the total suppression of the radiation
in a particular direction, $\alpha=\pi$ [$\sigma^H(\pi)=0$], and a
strong enhancement in the opposite direction, $\alpha=0$. The
complete three-dimensional radiation pattern resembles a cardioid or
apple-like shape, which is also azimuthally independent. Such a
radiation pattern of the Huygens source is potentially very useful
for various nanoantenna applications. However, while electric dipole
sources are widely used in optics, magnetic dipoles are less common.

First, we consider an electric dipole source placed in a close
proximity to a dielectric sphere [see Fig.~\ref{fig1}(a)]. As was mentioned above,
it can be analytically shown that high
permittivity dielectric nanoparticles exhibit strong magnetic
resonance in the visible range when the wavelength inside the
nanoparticle equals its diameter $\lambda/n_s\approx2R_s$~\cite{13},
where $n_S$ and $R_s$ are refractive index and radius of the
nanoparticle, respectively. There are many dielectric materials with
high enough real part of the permittivity and very low imaginary
part, indicating low dissipative losses. To name just a few, silicon
(Si, $\epsilon_{1}=16$), germanium (Ge, $\epsilon_{1}=20$), aluminum
antimonide (AlSb, $\epsilon_{1}=12$), aluminum arsenide (AlAs,
$\epsilon_{1}=10$), and other. In our study we concentrate on the
nanoparticles made of silicon, which support strong
magnetic resonance in the visible range for the radius varying from
$40$ nm to $80$ nm~\cite{9}.

For such a small radius compared to the wavelength $R_s<\lambda$,
the radiation pattern of the silicon nanoparticle in the far field
at the magnetic or electric resonances will resemble that of
magnetic or electric point-like dipole, respectively. Moreover, it
is even possible to introduce magnetic $\alpha^{m}$ and electric
$\alpha^{e}$ polarisabilities~\cite{9,52, Merchiers_07_PRA} based on
the Mie dipole scattering coefficients $b_1$ and $a_1$:

\begin{equation}\label{form1}
\alpha^{e}=\frac{6\pi a_{1} i}{k^{3}}, \; \;\alpha^{m}=\frac{6\pi
b_{1} i}{k^{3}}.
\end{equation}

Thus, the dielectric nanoparticle excited by the electric dipole
source at the magnetic resonance may result in the total far field
radiation pattern which is similar to that of the Huygens source.
Similar radiation patterns can be achieve in light scattering by a
magnetic particle when permeability equals permittivity
$\mu=\epsilon$, also known as Kerker's condition~\cite{Kerker:JOSA:1983}.
Our result suggests that even a dielectric
nonmagnetic nanoparticle can support two induced dipoles of equal
strength resulting in suppression of the radiation in the backward
direction. Thus, it can be considered as the simplest and efficient
optical nanoantenna with very good directivity.

In general, both polarisabilities $\alpha^{m}$ and $\alpha^{e}$ are
nonzero in the optical region ~\cite{9}. It is known that for a
dipole radiation in the far field the electric and magnetic
components should oscillate in phase to have nonzero energy flow. In
the near field the electric and magnetic components oscillate with
$\pi/2$ phase difference, thus, the averaged Poynting vector
vanishes, and a part of energy is stored in the vicinity of the
source. In the intermediate region, the phase between two components
varies form $\pi/2$ to 0. Placing a nanoparticle close to the
dipole source will change the phase difference between two
components, and, thus, will affect the amount of radiation form the near
field. In the case of plasmonic nanoparticles which exhibit electric
polarizability only, there is an abrupt phase change from 0 to $\pi$
in the vicinity of the localized surface plasmon resonance, which
makes it difficult to tune plasmonic nanoantennas for optimal
performance. The dependence of the scattering diagram on the
distance between the electric dipole source and metallic
nanoparticle was studied in Ref.~ \cite{Rolly_11_OptLett}. On
contrary, in the case of nanoparticles with both electric and
magnetic polarisabilities, it is possible to achieve more efficient
radiation from the near to far field zone, due to subtle phase
manipulation. {\em This is exactly the case of the dielectric
nanoparticles}.

Any antenna is characterized by two specific properties, directivity
$(D)$ and radiation efficiency $(\eta_{rad})$, defined as
~\cite{Novotny_10_NatPhot, 29}
\begin{equation}
D=\frac{4\pi}{P_{\mbox{rad}}}\mbox{Max}[p(\theta,\varphi)], \;\;
\eta_{rad}=\frac{P_{\mbox{rad}}}{P_{\mbox{rad}}+P_{\mbox{loss}}},
\label{directeff}
\end{equation}
where $P_{\mbox{rad}}$ and $P_{\mbox{loss}}$ are integrated radiated
and absorbed powers, respectively, $\theta$ and $\varphi$ are
spherical angles of standard spherical coordinate system, and $p(\theta,\varphi)$ is the radiated power in
the given direction $\theta$ and/or $\varphi$.
\begin{figure}[!t]
\centering \centerline{\includegraphics[width=0.65\textwidth]{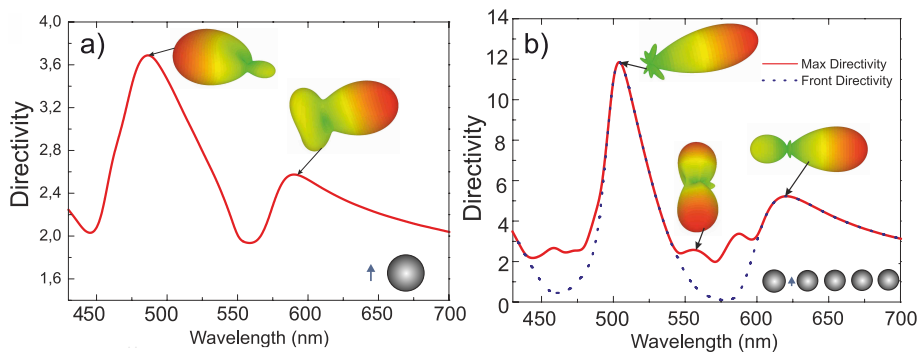}}
\caption{Wavelength dependence of the directivity of two types of
all-dielectric nanoantennas consisting of (a) single dielectric
nanoparticle of radius $R_d=70$ nm, and (b) Yagi-Uda like design
for the separation distance $D=70$ nm. Insert shows 3D radiation
pattern diagrams at particular wavelengths.~\cite{7}} \label{fig4}
\end{figure}
The directivity measures the power density of the antenna radiated
in the direction of its strongest emission, while Radiation
Efficiency measures the electrical losses that occur throughout the
antenna at a given wavelength. To calculate these quantities
numerically for the structures shown in Fig.~\ref{fig1}a, we employ
CST Microwave Studio. To get reliable results, we model the
electric dipole source by a Discrete Port coupled to two PEC
nanoparticles.

In Fig.~\ref{fig4}(a) we show the dependence of the directivity on
wavelength for a single dielectric nanoparticle excited by a
electric dipole source. Two inserts demonstrate 3D angular
distribution of the radiated pattern $p(\theta,\varphi)$
corresponding to the local maxima. In this case, the system radiates
predominantly to the forward direction at $\lambda=590$ nm, while in
another case, the radiation is predominantly in the backward
direction at $\lambda=480$ nm. In this case, the total electric
dipole moment of the sphere and point-like source and the magnetic
dipole moment of the sphere oscillate with the phase difference
$\mbox{arg}(\alpha^{m})-\mbox{arg}(\alpha^{e})=1.3\mbox{rad}$,
resulting in the destructive interference in the forward direction.
At the wavelength $\lambda = 590$ nm the total electric and magnetic
dipole moments oscillate in phase and produce Huygens-source-like
radiation pattern with the main lobe directed in the forward
direction.
\begin{figure}[!b]
\centering \centerline{\includegraphics[width=0.65\textwidth]{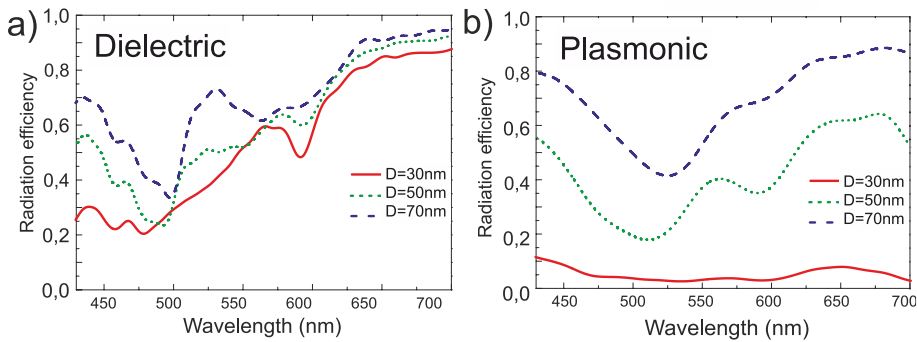}}
\caption{Radiation efficiencies of (a) dielectric (Si) and (b)
plasmonic (Ag) Yagi-Uda optical nanoantennas of the same geometrical
designs for various values of the separation distance $D$.~\cite{7}}
\label{fig5}
\end{figure}
By adding more elements to the silicon nanoparticle, we can enhance
the performance of all-dielectric nanoantennas. In particular, we
consider a dielectric analogue of the Yagi-Uda design (see
Fig.\ref{fig1}) consisting of four directors and one reflector. The
radii of the directors and the reflector are chosen to achieve the
maximal constructive interference in the forward direction along the
array. The optimal performance of the Yagi-Uda nanoantenna should be
expected when the radii of the directors correspond to the magnetic
resonance, and the radius of the reflector correspond to the
electric resonance at a given frequency, with the coupling between
the elements taken into account. Our particular design consists of
the directors with radii $R_d = 70$ nm and the reflector with the
radius $R_r = 75$ nm. In Fig.~\ref{fig4}(b) we plot the directivity
of all-dielectric Yagi-Uda nanoantenna vs. wavelength with the
separation distance $D = 70$ nm. Inserts demonstrate the 3D
radiation patterns at particular wavelengths. We achieve a strong
maximum at $\lambda = 500$ nm. The main lobe is extremely narrow
with the beam-width about $40^\circ$ and negligible backscattering.
The maximum does not correspond exactly to either magnetic or
electric resonances of a single dielectric sphere, which implies the
importance of the interaction between constitutive nanoparticles.

As the next step, we study the performance of the all-dielectric
nanoantennas for different separation distances $D$, and compare it
with {\em a plasmonic analogue} of the similar geometric design made
of silver nanoparticles. According to the results summarized in
Fig.~\ref{fig5}, the radiation efficiencies of both types of
nanoantennas {\em are nearly the same} for larger separation of
directors $D=70$ nm with the averaged value $70\%$. Although
dissipation losses of silicon are much smaller than those of silver,
the dielectric particle absorbs the EM energy by the whole spherical
volume, while the metallic particles absorb mostly at the surface.
As a result, there is no big difference in the overall performance
of these two types of nanoantennas for relatively large distances
between the elements. However, the difference becomes {\em very
strong} for smaller separations. The radiation efficiency of the
all-dielectric nanoantenna is insensitive to the separation distance
[see Fig.~\ref{fig5} (a)]. On contrary, the radiation efficiency
drops significantly for metallic nanoantennas [see Fig.~\ref{fig5}
(b)].

\begin{figure}[!t]
\centering \centerline{\includegraphics[width=0.65\textwidth]{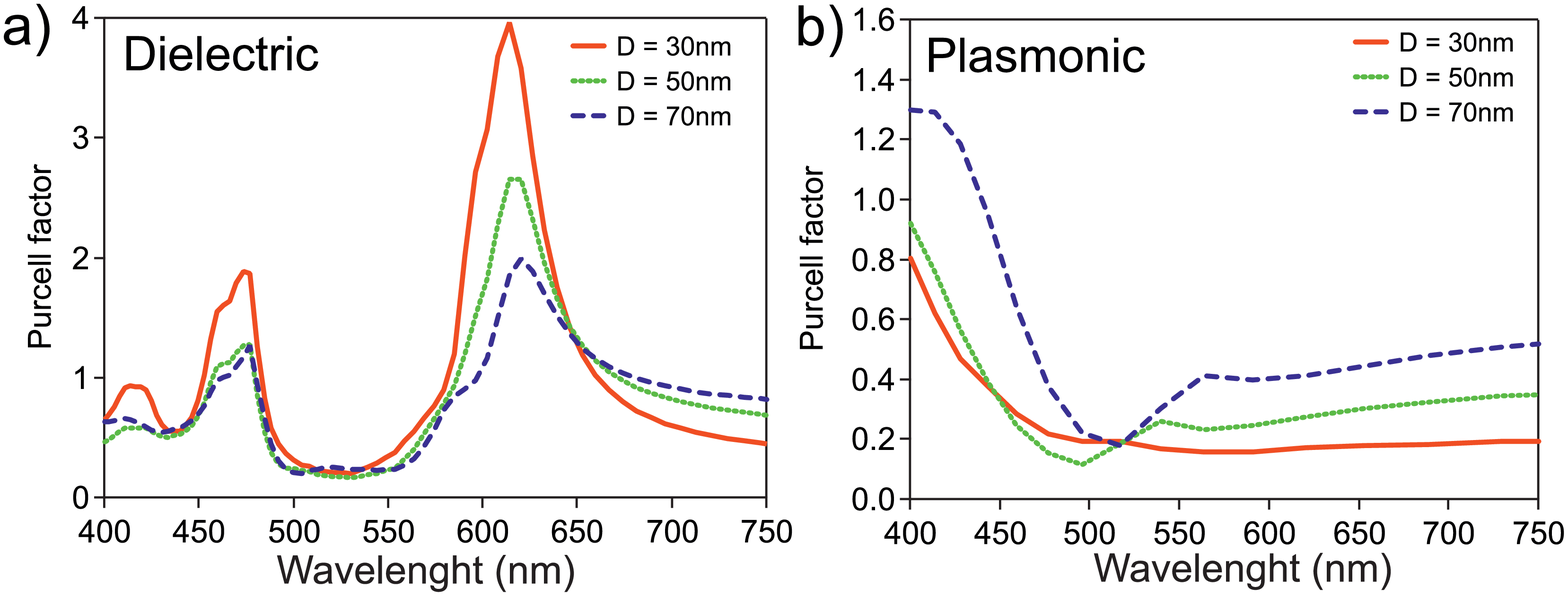}}
\caption{Purcell factor of all-dielectric Yagi-Uda nanoantenna vs
wavelength for various values of the separation distance
$D$.~\cite{7}} \label{fig6}
\end{figure}
Finally, we investigate the modification of the transition rate of a
quantum point-like source placed in the vicinity of dielectric
particles. For electric-dipole transitions and in the weak-coupling
regime, the normalised spontaneous decay rate $\Gamma/\Gamma_0$,
also known as Purcell factor, can be calculated classically as the
ratio of energy dissipation rates of an electric dipole
$P/P_0$~\cite{Novotny_Hecht_book}. Here, $\Gamma_0$ and $P_0$
correspond to transition rate of the quantum emitter and energy
dissipation rate of the electric dipole in free space~\cite{Chew}.
In the limit of the intrinsic quantum yield of the emitter close to
unity, both ratios become equal to each other
$\Gamma/\Gamma_0=P/P_0$, which allows us to calculate the Purcell
factor in the classical regime~\cite{Novotny_Hecht_book}. We have
calculated the Purcell factor by using both, numerical and
analytical approaches. Numerically, by using the CST Microwave
Studio we calculate the total radiated in the far-field and
dissipated into the particles powers and take the ratio of their sum
to the total power radiated by the electric dipole in free space.
Analytically, we employed the generalised multiparticle Mie
solution~\cite{Xu} adapted for the electric dipole
excitation~\cite{Dulkeith}. We verified that both approaches produce
similar results. In Fig.~\ref{fig6} we show calculated Purcell
factor of the all-dielectric Yagi-Uda nanoantenna vs. wavelength for
various separation distances. We observe that, by decreasing the
separation between the directors, the Purcell factor becomes
stronger near the magnetic dipole resonance. We can
notice that a plasmonic analogue of the same nanoantenna made of Ag exhibits low
Purcell factor less than one. Thus, such relatively high Purcell
factor can be employed for efficient photon extraction from molecules
placed near all-dielectric optical nanoantennas.

\subsection{Experimental verification of dielectric Yagi-Uda
nanoantenna}

There are exist some technological issues to reproduce an object of
the nanometer size with a high accuracy. For this
reason we have scaled the dimensions of the proposed optical all-dielectric Yagi-Uda nanoantenna to
the microwave frequency range while keeping all the material parameters in order to study the
microwave analogue of the nanoantenna experimentally. We use the design of the Yagi-Uda
antenna shown in Fig.~\ref{fig1}b. To mimic the silicon spheres in
microwave frequency range, we employ MgO-TiO$_{2}$ ceramic which is
characterized by dielectric constant of 16 and dielectric loss
factor of (1.12$-$1.17)10$^{-4}$ measured at frequency 9-12 GHz
~\cite{Nenasheva_03}. As a source, we use a half-wavelength
vibrator. We study experimentally both the radiation pattern and
directivity of the antenna.

We set the radius of the reflector equal to $R_{r}=5$ mm. The
frequencies of the electric and magnetic Mie resonances of the
sphere calculated with the help of Eq.~(\ref{form1}) are 10.2 GHz
and 7 GHz, respectively. The radius of the directors is $R_{d}=4$
mm. In this case, the frequencies of the electric and magnetic Mie
resonances are 12.5 GHz and 9 GHz. As a source, we model a
half-wavelength vibrator with the total length of $L_{v}=19.8$ mm
and diameter of $D_{v}=2.2$ mm. The distances between the reflector, directors, and vibrator
have been adjusted by numerical simulations.
We achieve an effective suppression of the back and minor lobes, and
the narrow major lobe (of about 40$^{\circ}$) of the antenna when
the distance between the director's surface as well as the distance
between vibrator center and the first director surface are 1.5 mm;
the distance between the surface of the reflector and vibrator
centre is 1.1 mm.

\begin{figure}
\centering \epsfig{file=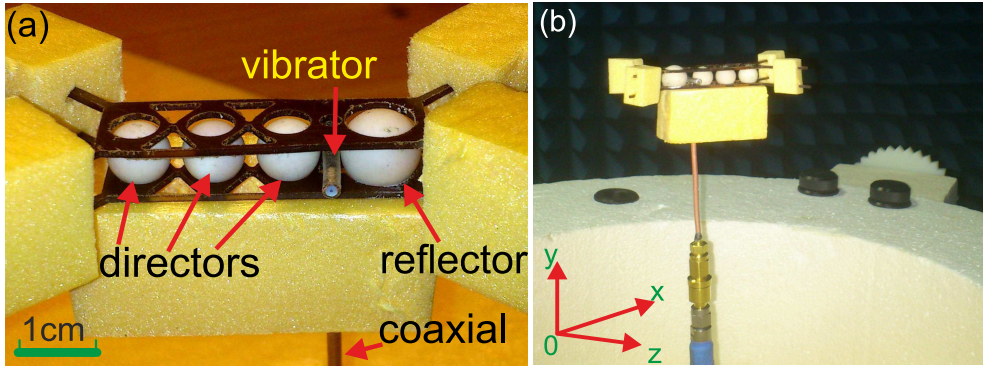,width=0.7\textwidth}
\caption{Photographs of the all-dielectric Yagi-Uda microwave
antenna. (a) Detailed view of the antenna placed in a holder. (b)
Antenna placed in an anechoic chamber; the coordinate $z$ is
directed along the vibrator axis; the coordinate $y$ is directed
along the antenna axis.~\cite{8}} \label{fig7}
\end{figure}
Figures~\ref{fig7}(a,b) show the photographs of the fabricated
all-dielectric Yagi-Uda antenna. The reflector and directors are
made of MgO-TiO$_{2}$ ceramic with accuracy of $\pm$0.05 mm. To
fasten together the elements of the antenna and vibrator, we use a
special holder made of a thin dielectric substrate with dielectric
permittivity close to 1 [being shown in Fig.~\ref{fig7}(a)].
Styrofoam material with the dielectric permittivity of 1 is used to
fix the antenna in the azimuthal-rotation unit [see
Fig.~\ref{fig7}(b)]. To feed the vibrator, we employ a coaxial cable
that is connected to an Agilent PNA E8362C vector network analyzer.

\begin{figure}[!b]
\centering \epsfig{file=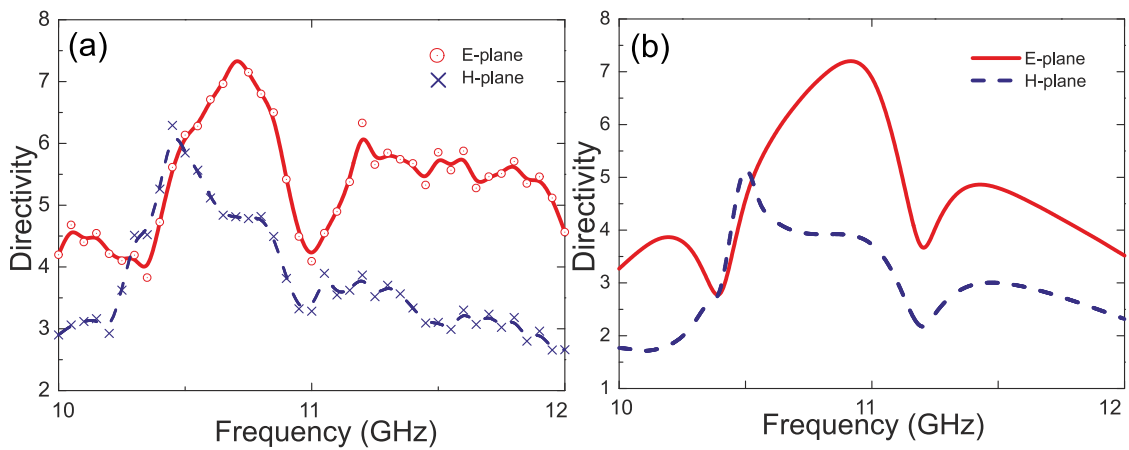,width=0.8\textwidth}
\caption{(a )Experimentally measured and (b) numerically calculated
antenna's directivity in both $E$- and $H$-planes.~\cite{8}}
\label{fig8}
\end{figure}

\begin{figure}[!t]
\centering \epsfig{file=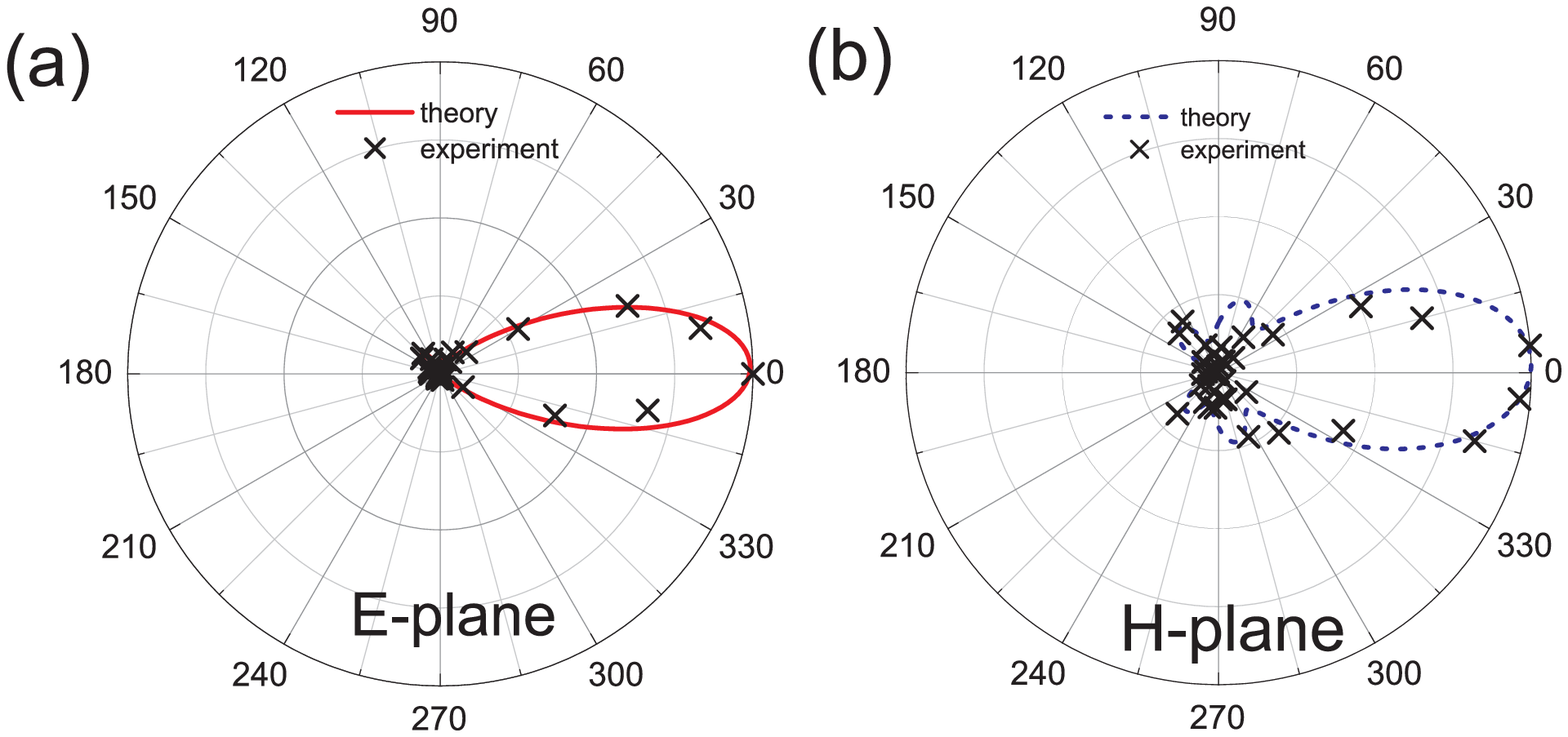,width=0.7\textwidth}
\caption{Radiation pattern of the antenna in (a) $E$-plane and (b)
$H$-plane at the frequency 10.7 GHz. Solid lines show the result of
numerical simulations in CST; the crosses correspond to the
experimental data.~\cite{8}} \label{fig9}
\end{figure}

Any antenna is characterized by the total directivity
(\ref{directeff}). Sometimes it is not possible to determine the
value of the total directivity experimentally due to difficulties to
measure the total radiated power $P_{\mbox{rad}}$. In this case, it
is convenient to use directivity in the planes where electric field
$\mathbf{E}$ and magnetic field $\mathbf{H}$ oscillate in the far
field. For our coordinates the directivity in the evaluation plane
(E-plane) and the azimuthal plane (H-plane) can be expressed as:
\begin{equation}
D_{E}=\left.\frac{2\pi\mbox{Max}[p(\theta)]}{\int_{0}^{2\pi}p(\theta)d\theta}\right|_{\varphi=0},\;\;
D_{H}=\left.\frac{2\pi\mbox{Max}[p(\varphi)]}{\int_{0}^{2\pi}p(\varphi)d\varphi}\right|_{\theta=\pi/2}.
\label{form2}
\end{equation}
Equations~(\ref{form2}) are multiplied by $2\pi$ because of the
integration in the denominator is performed only for one coordinate
while the second coordinate is fixed.

To extract the antenna directivity in the $E$- and $H$-planes from
the experimental data, we measure the radiated power by the antenna
in the frequency range from 10 GHz to 12 GHz with a step of 50 MHz.
Then, by employing Eq.~(\ref{form2}) we calculate the directivity at
each frequency. The results are presented in Fig.~\ref{fig8}a. To
estimate the performance of the all-dielectric Yagi-Uda antenna at
microwaves, we simulate numerically the antenna's response by
employing the CST Microwave Studio. We observe excellent agreement
between numerical results of Fig.~\ref{fig8}b and measured
experimental data. However, we notice a small frequency shift of the
measured directivity (approx. $2\%$) in comparison with the
numerical results. This discrepancies can be explained by the effect
of the antenna holders in the experiment, not included into the
numerical simulation.

The antenna radiation patterns in the far field (at the distance
$\simeq3$ m) are measured in an anechoic chamber by a horn antenna
and rotating table. The measured radiation patterns of the antenna
in $E$- and $H$-planes at the frequency 10.7 GHz are shown in
Fig.~\ref{fig9}. The measured characteristics agree very well with
the numerical results. A small disagreement can be explained by the
presence of the antenna holder which influence was not taken into
account in our numerical simulations.

\section{All--dielectric superdirective optical
nanoantenna}\label{SupDirect}

For optical wireless circuits on a chip, nanoantennas are required
to be both highly directive and compact~\cite{4, Halas, 26,
Landesa}. In nanophotonics, directivity has been achieved for
arrayed plasmonic antennas utilizing the Yagi-Uda
design~\cite{Novotny_10_NatPhot, Giessen11, 26,KrasnokOE,Chew12},
large dielectric spheres~\cite{32}, and metascreen
antennas~\cite{Eleftheriades}. Though individual elements of these
arrays are optically small, the overall size of the radiating
systems is larger than the radiation wavelength $\lambda$. In
addition, small plasmonic nanoantennas possess weak directivity
close to the directivity of a point dipole~\cite{26, 34,Rodriguez}.

As was discussed above, it was suggested theoretically and experimentally to
employ magnetic resonances of high-index dielectric nanoparticles
for enhancing the nanoantenna directivity~\cite{58, 7, 8,
Krasnok_12AIP, SPIEKrasnok, Noskov12Arxiv, nanoBook, KrasnokOE,
BonodOE}. High-permittivity nanoparticles can have nearly resonant
balanced electric and magnetic dipole responses. This balance of the
electric and magnetic dipoles oscillating with the same phase allows
the practical realization of the Huygens source, an elementary
emitting system with a cardioid pattern~\cite{29, Krasnok_11, 8,
KrasnokOE} and with the directivity larger than 3.5. Importantly, a
possibility to excite magnetic resonances leads to the improved
nanoantenna directional properties without a significant increase of
its size.

Superdirectivity has been already discussed for radio-frequency
antennas, and it is defined as directivity of an electrically small
radiating system that significantly exceeds (at least in 3 times)
directivity of an electric dipole~\cite{29, Hansen, 38}. In that
sense, the Huygens source is not superdirective. In the antenna
literature, superdirectivity is claimed to be achievable only in
antenna arrays by the price of ultimately narrow frequency range and
by employing very precise phase shifters (see, e.g., Ref.~\cite{29,
Hansen, 38}). Therefore, superdirective antennas, though very
desirable for many applications such as space communications and
radioastronomy, were never demonstrated and implemented for
practical applications.

Superdirectivity was predicted theoretically for an antenna
system~\cite{Eleftheriades} where some phase shifts were required
between radiating elements to achieve complex shapes of the elements
of a radiating system which operates as an antenna array. In this
paper, we employ the properties of subwavelength particles excited
by an inhomogeneous field with higher-order magnetic multipoles. We
consider a subwavelength dielectric nanoantenna (with the size of
0.4 wavelength) with a notch resonator excited by a point-like
emitter located in the notch. The notch transforms the energy of the
generated magneto-dipole Mie resonance into high-order multipole
moments, where the magnetic multipoles dominate. This system is
resonantly scattering i.e. it is very different from dielectric
lenses and usual dielectric cavities which are large compared to the
wavelength. Another important feature of the notched resonator is its huge sensitivity of the radiation direction to a spatial position of the emitter. This property leads to a strong beam steering effect
and subwavelength sensitivity of the radiation direction to the
source location. The proposed design of superdirective nanoantennas
may also be useful for collecting single-source radiation,
monitoring quantum objects states, and nanoscale microscopy.
\begin{figure}[!b]
\centerline{\includegraphics[width=.8\textwidth]{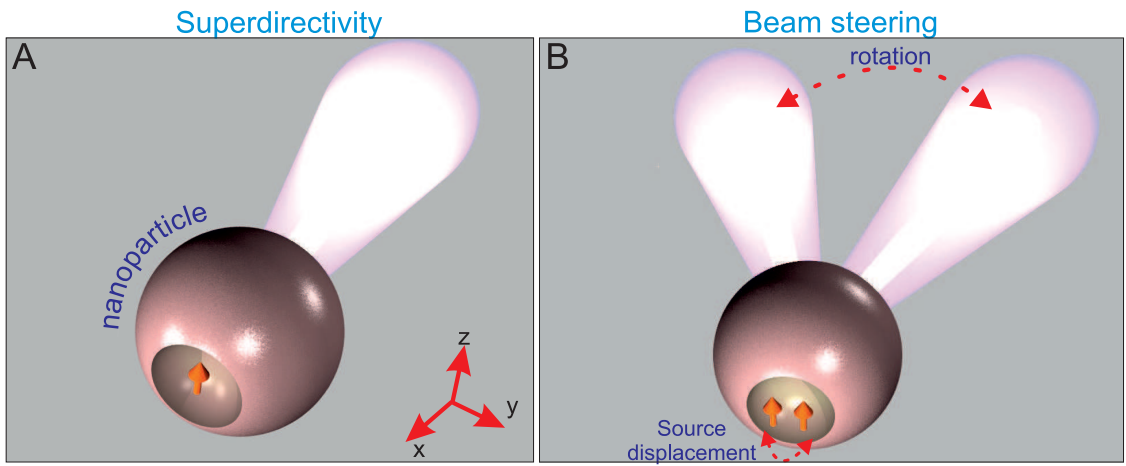}}
\caption{({\bf A}) Geometry of an all-dielectric superdirective
nanoantenna excited by a point-like dipole. ({\bf B}) Concept of the
beam steering effect at the nanoscale.} \label{geometr}
\end{figure}
In order to achieve superdirectivity, we should generate
subwavelength spatial oscillations of the radiating
currents~\cite{29, Hansen, 38}. Then, near fields of the antenna
become strongly inhomogeneous, and the near-field zone expands
farther than that of a point dipole. The effective antenna aperture can be defined as
$\mbox{S}=\mbox{D}_{\mbox{max}}\lambda^2/(4\pi)$,
where the maximum of directivity $\mbox{D}_{\mbox{max}}=4\pi
\mbox{P}_{\mbox{max}}/\mbox{P}_{\mbox{tot}}$, $\lambda$ is the wavelength in free space in our case,
$\mbox{P}_{\mbox{max}}$ and $\mbox{P}_{\mbox{tot}}$ are respectively the maximum power in the direction of the radiation pattern and the total radiation power. By normalizing the effective aperture
$\mbox{S}$ by the geometric aperture for a spherical antenna
$\mbox{S}_0=\pi \mbox{R}_{\mbox{s}}^2$, we obtain the definition of superdirectivity~\cite{Hansen, 29}:

\begin{eqnarray}
\mbox{S}_{n}=\frac{\mbox{D}_{\mbox{max}}\lambda^2}{4\pi^2\mbox{R}_{\mbox{s}}^2}\gg1
\end{eqnarray}

Practically, the value $\mbox{S}_{n}=4\dots 5$ is sufficient for
superdirectivity of a sphere. In this work, maximum of 6.5 for
$\mbox{S}_{n}$ is predicted theoretically for the optical frequency
range, and the value of 5.9 is demonstrated experimentally for the
microwave frequency range.

\subsection{Concept of all--dielectric superdirective optical
nanoantennas}

\textit{Here we demonstrate a possibility to create a superdirective
nanoantenna without hypothetic metamaterials and plasmonic arrays.}
We consider a silicon nanoparticle, taking into account the
frequency dispersion of the dielectric permittivity~\cite{Palik}.
The radius of the silicon sphere is equal in our example to
$\mbox{R}_{\mbox{s}}=90$ nm. For a simple sphere under rather
homogeneous (e.g. plane-wave) excitation, only electric and magnetic
dipoles can be resonantly excited while the contribution of
higher-order multipoles is negligible in the visible~\cite{KrasnokOE}. Making a
notch in the sphere breaks the symmetry and increases the
contribution of higher-order multipoles into scattering even if the
sphere is still excited homogeneously. Further, placing a
nanoemitter (e.g. a quantum dot) inside the notch, as shown in
Fig.~\ref{geometr} we create the conditions for the resonant
excitation of multipoles: the field exciting the resonator is now
spatially very non-uniform as well as the field of a set of
multipoles. In principle, the notched particle operating as a nanoantenna can be performed by different semiconductor materials and have various shapes -- spherical, ellipsoidal, cubic, conical,
as well as the notch. However, in this work, the particle is a
silicon sphere and the notch has the shape of a hemisphere with a
radius $\mbox{R}_{\mbox{n}}< \mbox{R}_{\mbox{s}}$. The emitter is
modeled as a point-like dipole and it is shown in
Fig.~\ref{geometr} by a red arrow.

It is important to mention that our approach is seemingly close to
the idea of references~\cite{Sveta1, Wang} where a small notch on a
surface of a semiconductor microlaser was used to achieve higher
emission directivity by modifying the field distribution inside the
resonator~\cite{Scully}. An important difference between those
earlier studies and our work is that the design discussed earlier is
not optically small and the directive emission is not related to
superdirectivity. In our case, the nanoparticle is much smaller than
the wavelength, and our design allows superdirectivity. For the same
reason our nanoantenna is not dielectric~\cite{Kim09, Lukyanchuk11}
or Luneburg~\cite{Lipson11, Leonhardt11} lenses. For example,
immersion lenses~\cite{Rigneault08, Quake10, Hanson11, Wrachtrup10} are
the smallest from known dielectric lenses, characterized by the
large size 1-2 $\mu$m in optical frequency range. The working methodology of such lenses is to collect a radiation by large geometric aperture $\mbox{S}$, while $\mbox{S}_{n}\simeq1$. \textit{Our approach
demonstrates that the subwavelength system, with small geometric
aperture, can have high directing power because of an increase of
the effective aperture.} Moreover, there are articles (see. references
~\cite{Chew, Klimov}) where the transition rates of atoms inside and
outside big dielectric spheres with low dielectric constant
(approximately 2), were studied.

\begin{figure}[!b]
\centerline{\includegraphics[width=.8\textwidth]{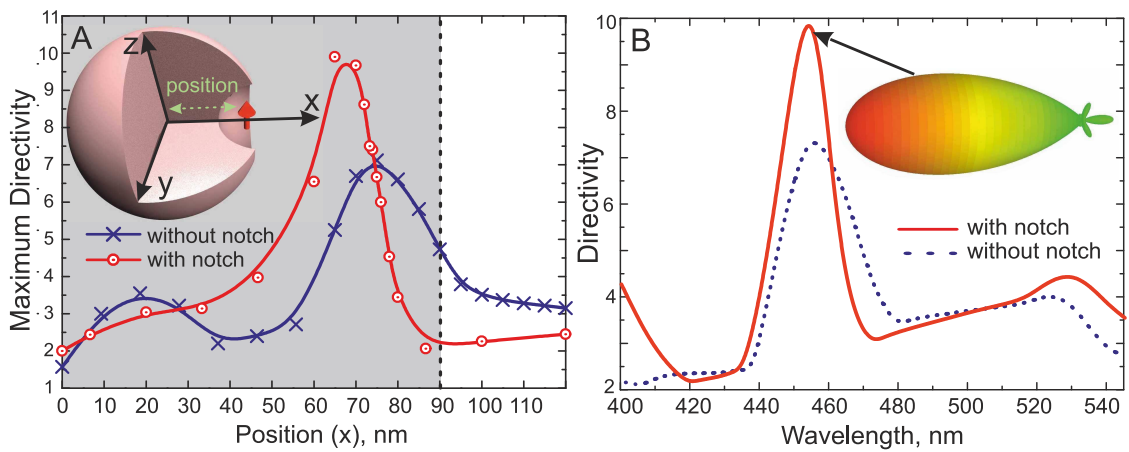}}
\caption{({\bf A}) Maximum of directivity depending on the position
of the emitter ($\lambda=455$ nm) in the case of a sphere with and
without notch. Vertical dashed line marks the particle radius
centered at the coordinate system. ({\bf B}) Directivity dependence
on the radiation wavelength. The inset shows three-dimensional
radiation pattern of the structure ($\mbox{R}_{\mbox{s}}=90$~nm and
$\mbox{R}_{\mbox{n}}=40$~nm).} \label{direct}
\end{figure}
First, we consider a particle without a notch but excited inhomogeneously by an emitter point.
To study the problem numerically, we employed the simulation software CST Microwave
Studio. Image Fig.~\ref{direct}A shows the dependence of the
maximum directivity $\mbox{D}_{\mbox{max}}$ on the position of the
source in the case of a sphere $\mbox{R}_{\mbox{s}}=90$~nm without a
notch, at the wavelength $\lambda=455$~nm (blue curve with crosses).
This dependence has the maximum ($\mbox{D}_{\mbox{max}}=7.1$) when
the emitter is placed inside the particle at the distance 20~nm from
its surface. The analysis shows that in this case the electric field
distribution inside a particle corresponds to the noticeable
excitation of higher-order multipole modes not achievable with the
homogeneous excitation.

Furthermore, the amplitudes of high-order multipoles are
significantly enhanced with a small notch around the emitter,
as it is shown in Fig.~\ref{geometr}. This geometry transforms it
into a resonator with high-order multipole moments. In this example
the center of the notch is on the nanosphere's surface. The optimal
radius of the notch (for maximal directivity) is $\mbox{R}_{\mbox{n}}=40$ nm. In Fig.~\ref{direct}A the
extrapolation red curve with circles, corresponding to simulation
results, shows the maximal directivity versus the location of the
emitter at the wavelength 455~nm. The Fig.~\ref{direct}B shows the directivity versus $\lambda$ with and without a notch, it exhibits a maximum of 10 for the directivity at 455 nm. The inset shows the
three-dimensional radiation pattern of the structure at
$\lambda=$455 nm. This pattern has an angular width (at the level of
3 dB) of the main lobe equal to $40^{\circ}$. This value of
directivity corresponds to the normalized effective aperture
$\mbox{S}_{n}=6.5$.

\begin{figure}[!t]
\centerline{\includegraphics[width=.6\textwidth]{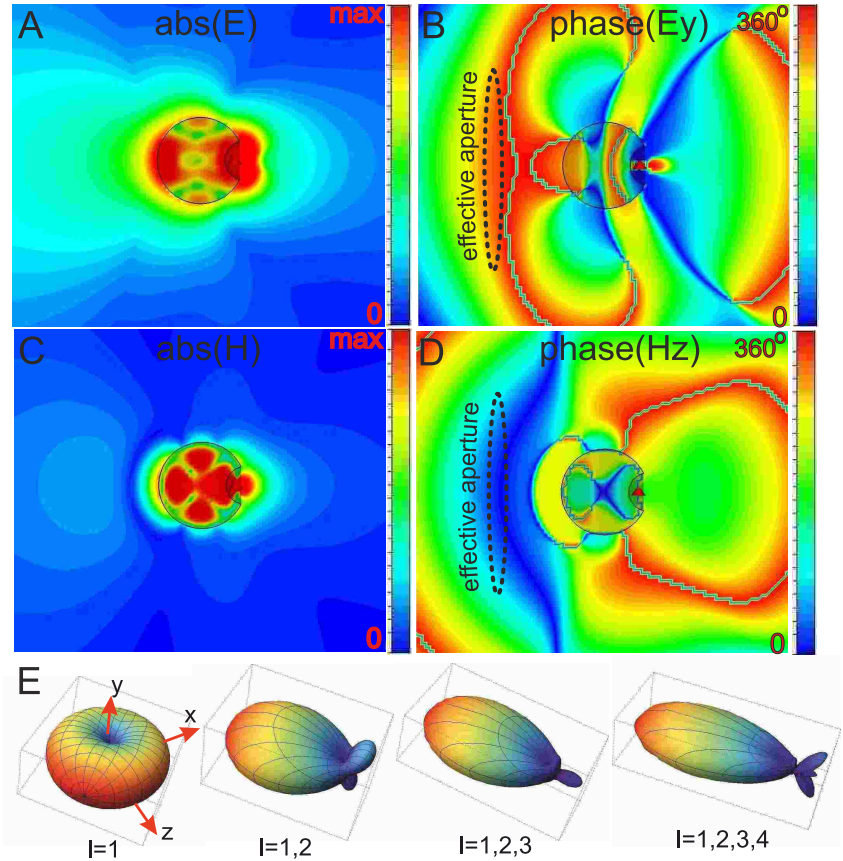}}
\caption{Distribution of ({\bf A}) absolute values and ({\bf B})
phases of the electric field ({\bf C} and {\bf D} for magnetic
field, respectively) of the all-dielectric superdirective
nanoantenna with source in the center of notch, at the wavelength
$\lambda=455$ nm. ({\bf E}) Dependence of the radiation pattern of
all-dielectric superdirective nanoantenna on the number of taken
into account multipoles. Dipole like source located along the $z$
axis.} \label{fieldPatt}
\end{figure}

Figures Fig.~\ref{fieldPatt}A and B show the distribution of the absolute values and phases of the internal electric field in the vicinity of the nanoantenna. Electric and magnetic fields inside the particle are strongly inhomogeneous at $\lambda=$455~nm i.e. in the
regime of the maximal directivity. In this regime, the
internal area where the electric field oscillates with approximately
the same phase turns out to be maximal. This area is located near
the back side of the spherical particle, as can be seen in figure
Fig.~\ref{fieldPatt}B,D. In other words, the effective near zone
of the nanoantenna is maximal in the superdirective regime.

Usually, high directivity of plasmonic nanoantennas is achieved by
the excitation of higher \textit{electrical} multipole moments in
plasmonic nanoparticles~\cite{Rolly_11_OptLett, Kall09, Pakizeh12}
or for core-shell resonators consisting of a plasmonic material and
a hypothetic metamaterial which would demonstrate the extreme
material properties in the nanoscale~\cite{Alu}. Although, the
values of directivity achieved for such nanoantennas do not allow
superdirectivity, these studies stress the importance of higher
multipoles for the antenna directivity.

Next, we demonstrate how to find multipole modes excited in the
all-dielectric superdirective nanoantenna which are responsible for
its enhanced directivity. We expand the exactly simulated internal
field, producing the polarization currents in the nanoparticle, into
multipole moments following to~\cite{Jackson}. The expansion is a
series of vector spherical harmonics with the coefficients $a_{E}(l,m)$ and
$a_{M}(l,m)$, which characterize the electrical and magnetic
multipole moments~\cite{Jackson}:

\begin{eqnarray}
a_{E}(l,m)&=&\frac{4\pi
k^2}{i\sqrt{l(l+1)}}\int{Y_{lm}^{*}\left[\rho\frac{\partial}{\partial
r}[rj_{l}(kr)]+
\frac{ik}{c}(\mathbf{r}\cdot\mathbf{j})j_{l}(kr)\right]}d^{3}x, \nonumber \\
a_{M}(l,m)&=&\frac{4\pi
k^2}{i\sqrt{l(l+1)}}\int{Y_{lm}^{*}\mbox{div}\left(\frac{\mathbf{r}\times
\mathbf{j}}{c}\right)j_{l}(kr)}d^{3}x, \label{moments}
\end{eqnarray}
where $\rho=1/(4\pi)\mbox{div}(\mathbf{\mbox{E}})$ and
$\mathbf{j}=c/(4\pi)(\mbox{rot}(\mathbf{\mbox{H}})+ik\mathbf{\mbox{E}})$
are densities of the \textit{total} electrical charges and currents
that can be easily expressed through the internal electric
$\mathbf{\mbox{E}}$ and magnetic $\mathbf{\mbox{H}}$ fields of the
sphere, $Y_{lm}$ are the spherical harmonics of the orders $(l>0$ and $0\ge
|m|\le l)$, $k=2\pi/\lambda$, $j_{l}(kr)$ are the $l$-order spherical Bessel
function and $c$ is the speed of light. Coefficients
$a_{E}(l,m)$ and $a_{M}(l,m)$ determine the electric and magnetic
mutipole moments, namely dipole at $l=1$, quadrupole at $l=2$,
octupole at $l=3$ etc.

The multipole coefficients determine not only the mode structure of
the internal field but also the angular distribution of the
radiation. In particular, in the far field zone electric and
magnetic fields of $l$-order multipole depend on the distance $r$ as
~\cite{Jackson} $\sim(-1)^{i+1}\frac{\exp(ikr)}{kr}$ and expression
for the angular distribution of the radiation power can be written
as follows:

\begin{eqnarray}
\frac{\mbox{d}P(\theta,\varphi)}{\mbox{d}\Omega}&=&\frac{c}{8\pi
k^2}\left|\sum_{l,m}(-i)^{l+1}[a_{E}(l,m)\mathbf{X}_{lm}\times\mathbf{n}+a_{M}(l,m)\mathbf{X}_{lm}]\right|^2,\nonumber\\
\mathbf{X}_{lm}(\theta,\varphi)&=&\frac{1}{\sqrt{l(l+1)}}
\begin{bmatrix}
A_{l,m}^{-}Y_{l,m+1}+A_{l,m}^{+}Y_{l,m-1}\\
-iA_{l,m}^{-}Y_{l,m+1}+iA_{l,m}^{+}Y_{l,m-1}\\
mY_{l,m}
\end{bmatrix},
\label{fpattern}
\end{eqnarray}

\begin{figure}[!b]
\centerline{\includegraphics[width=.6\textwidth]{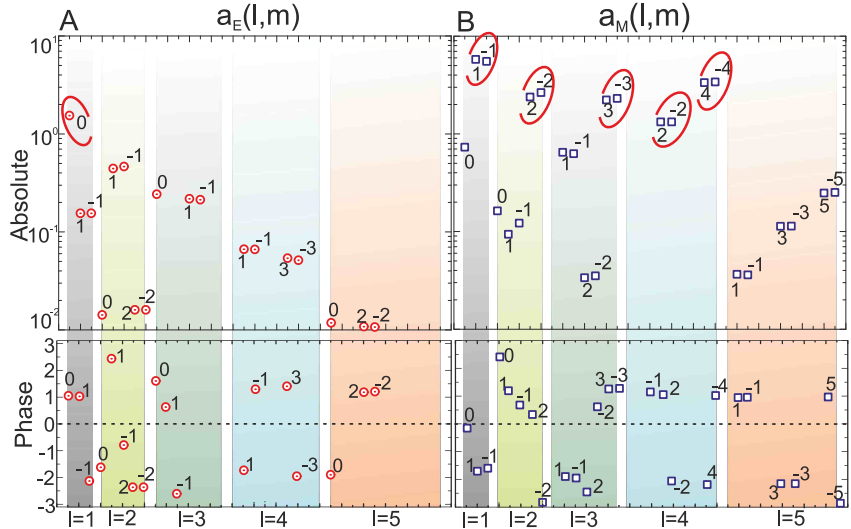}}
\caption{Absolute values and phases of ({\bf A}) electric and ({\bf
B}) magnetic multipole moments that provide the main contribution of
the radiation of all-dielectric superdirective optical nanoantenna
at the wavelength 455~nm. Multipole coefficients providing the
largest contribution to the antenna direction are highlighted by red
circles.} \label{allmoments1}
\end{figure}

where $A_{l,m}^{\pm}=(1/2)\sqrt{(l\pm m)(l\mp m+1)}$,
$\mbox{d}\Omega=\mbox{sin}(\theta)\mbox{d}\theta\mbox{d}\varphi$ is
the solid angle element in spherical coordinates and $\mathbf{n}$ -
unit vector of the observation point. All coefficients $a_E(l,m)$
and $a_M(l,m)$ give the same contribution to the radiation, if they
have the same values. Since higher-order multipoles for optically
small systems have usually negligibly small amplitudes compared to
$a_E(1,m)$ and $a_M(1,m)$, they are, as a rule, not considered.

The amplitudes of multipole moments, are found by using the
expressions \eqref{moments} for electric and magnetic fields
distribution Fig.~\ref{fieldPatt}A-D are shown in
Fig.~\ref{allmoments1}, where we observe strong excitation of
$a_{E}(1,0)$, $a_{M}(1,1)$, $a_{M}(1,-1)$, $a_{M}(2,2)$,
$a_{M}(2,-2)$, $a_{M}(3,3)$, $a_{M}(3,-3)$, $a_{M}(4,2)$,
$a_{M}(4,-2)$, $a_{M}(4,4)$ and $a_{M}(4,-4)$. These multipole
moments determine the angular pattern of the antenna. All other ones
give a negligible contribution. Absolute values of all magnetic
moments are larger than those of the electric moments in the
corresponding multipole orders, and the effective spectrum of
magnetic multipoles is also broader than the one of the electric moments. Thus, the operation of the antenna is mainly determined by
the magnetic multipole response. Absolute values of multipole
coefficients $a_{M}(l,\pm |m|)$ of the same order $l$ are
practically equivalent. However, the phase of some coefficients are
different. Therefore, the modes with $+|m|$ and $-|m|$ form a strong
anisotropy of the forward--backward directions that results in the
unidirectional radiation.

We have performed the transformation of multipole coefficients into
an angular distribution of radiation in accordance to
\eqref{fpattern} by using distribution of the electric and magnetic
fields Fig.~\ref{fieldPatt}A-D and determined the relative
contribution of each order $l$. Fig.~\ref{fieldPatt}E shows how the
directivity grows versus the spectrum of multipoles with equivalent
amplitudes. The right panel of Fig.~\ref{fieldPatt}E nearly
corresponds to the inset in Fig.~\ref{direct} that fits to the
results shown in Fig.~\ref{fieldPatt}E.

Generally, the superdirectivity effect is accompanied by a
significant increase of the effective near field zone of the antenna
compared to the one of a point dipole for which the near zone radius is
equal to $\lambda/2\pi$. In the optical frequency range this effect is
especially important, considering the crucial role of the near
fields at the nanoscale.

Usually, the superdirectivity regime corresponds to a strong
increase of dissipative losses~\cite{29}. Radiation efficiency of
the nanoantenna is determined by
$\eta_{\mbox{rad}}=\mbox{P}_{\mbox{rad}}/\mbox{P}_{\mbox{in}}$,
where $\mbox{P}_{\mbox{in}}$ is the accepted input power of the nanoantenna. However, the multipole moments excited in our nanoantenna are mainly of magnetic type that leads to a strong
increase of the near magnetic field that dominates over the electric
one. Since the dielectric material does not dissipate the magnetic
energy, the effect of superdirectivity does not lead to a so large
increase of losses in our nanoantenna as it would be in the case of
dominating electric multipoles. However, since the electric near
field is nonzero the losses are not negligible. At wavelengths
440-460 nm (blue light) the directivity achieves 10 but the
radiation efficiency is less than 0.1 (see
[Fig.~\ref{directEffDisper1})]. This is because silicon has very
high losses in this range~\cite{Palik}. Peak of directivity is
shifted to longer wavelengths with the increase of the nanoantenna size. For the design parameters corresponding to the operation wavelength 630 nm (red light) the calculated value of
radiation efficiency is as high as 0.5, with nearly same directivity
close to 10. In the infrared range, there are high dielectric
permittivity materials with even lower losses. In principle, the
proposed superdirectivity effect is not achieved by price of
increased losses, and this is an important advantage compared to
known superdirective radio-frequency antenna arrays~\cite{29} and
compared to their possible optical analogues -- arrays of plasmonic
nanoantennas.

\begin{figure}[!b]
\centerline{\includegraphics[width=.6\textwidth]{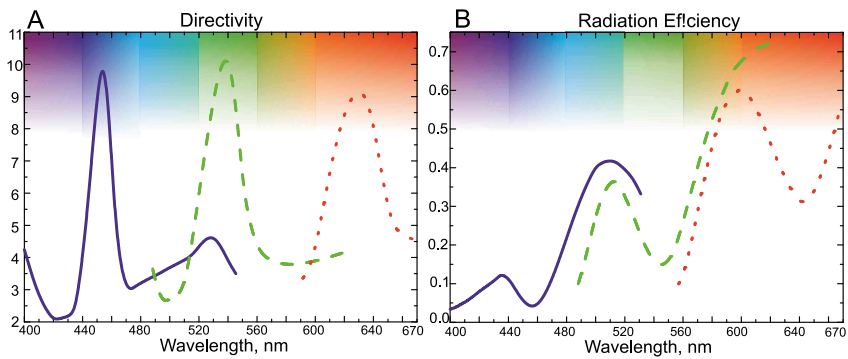}}
\caption{Dependence of directivity ({\bf A}) and radiation
efficiency ({\bf B}) on the size of nanoantenna. Here, the blue
solid lines corresponds to the geometry -- $\mbox{R}_{\mbox{s}}=90$
nm, $\mbox{R}_{\mbox{n}}=40$ nm, the green dashed curves --
$\mbox{R}_{\mbox{s}}=120$ nm, $\mbox{R}_{\mbox{n}}=55$ nm and red
point curves -- $\mbox{R}_{\mbox{s}}=150$ nm,
$\mbox{R}_{\mbox{n}}=65$ nm. Growth of the nanoantenna efficiency
due to the reduction of dissipative losses in silicon with
increasing of wavelength.} \label{directEffDisper1}
\end{figure}

\subsection{Steering of light at the nanoscale}

\begin{figure}[!t]
\centerline{\includegraphics[width=.6\textwidth]{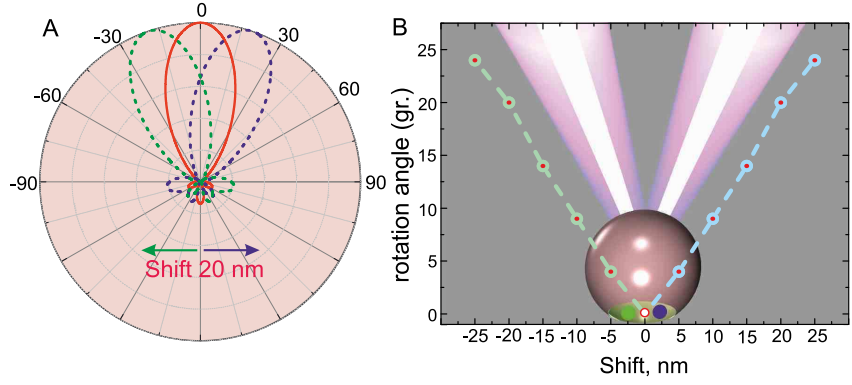}}
\caption{The rotation effect of the main beam radiation pattern,
with subwavelength displacement of emitter inside the notch. ({\bf
A}) The radiation patterns of the antenna with the source in center
(solid line) and the rotation of the beam radiation pattern for the
20 nm left/right offset (dashed lines). ({\bf B}) Dependence of the
rotation angle on the source offset.} \label{shiftEffect}
\end{figure}

Here we examine the response of the nanoantenna to subwavelength
displacements of the emitter. Displacement in the plane
perpendicular to the axial symmetry of antenna (i.e. along the $y$
axis) leads to the rotation of the beam \emph{without damaging the
superdirectivity}. Fig.\ref{shiftEffect}A shows the
radiation patterns of the antenna with the source in center (solid
line) and the rotation of the beam for the 20 nm left/right offset
(dashed lines). Shifting of the source in the right side leads to the rotation
of pattern to the left, and vice versa. The angle of the beam
rotation is equal to 20 degrees, that is essential and available to
experimental observations. The result depends on the geometry of the
notch. For a hemispherical notch, the dependence of the rotation
angle on the displacement is presented in Fig.\ref{shiftEffect}B.

\begin{figure}[!b]
\centerline{\includegraphics[width=.6\textwidth]{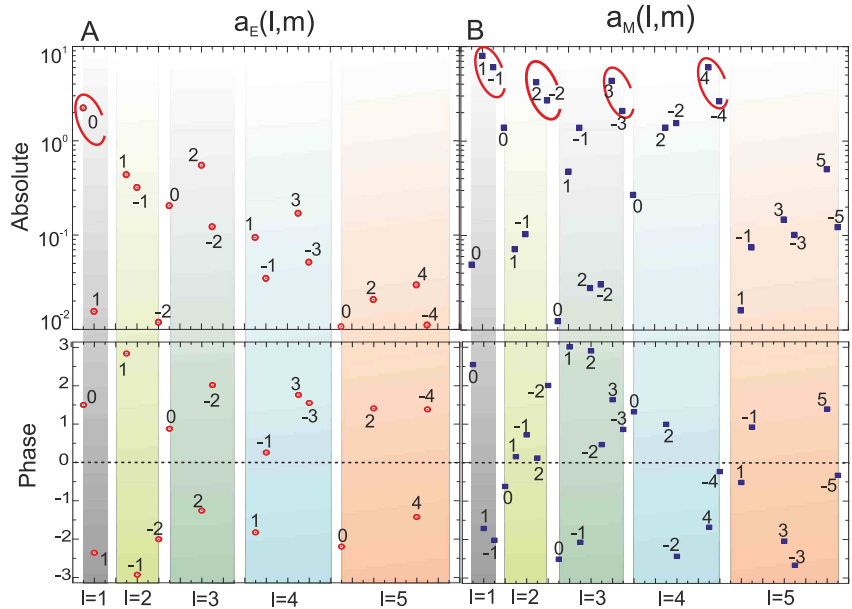}}
\caption{Absolute values and phases of ({\bf A}) electric and ({\bf
B}) magnetic multipole moments that provide the main contribution to
the radiation of all-dielectric superdirective optical nanoantenna
in case of asymmetrical location of source at the wavelength 455~nm.
Coefficients that give the largest contribution to the antenna
directivity are highlighted by red circles.} \label{shiftmagnet}
\end{figure}

To interpret the beam steering effect, we can consider the result of field expansion to electric and magnetic multipoles, as shown in Fig.\ref{shiftmagnet}. In the case of asymmetrical location (the 20 nm left offset) of the source in the notch absolute values of $a_{M}(l,\pm |m|)$ are different. This means that the mode $a_{M}(l,+|m|)$ is excited more strongly than $a_{M}(l,-|m|)$, or vice versa, that depends on direction of displacement. The effect of superdirectivity remains even with an offset of the source until to the edge of the notch. Small displacements of the source along $x$ and $z$ do not lead to the rotation of the pattern.

Instead of the movement of a single quantum dot one we can have
the emission of two or more quantum dots located near the edges of
the notch. In this case, the dynamics of their spontaneous decay
will be well displayed in the angular distribution of the radiation.
This can be useful for quantum information processing and for
biomedical applications.

Beam steering effect described above is similar to the effect of
beam rotation in hyperlens~\cite{Zhang07, Narimanov06, LiuReview12},
where the displacement of a point-like source leads to a change of
the angular distribution of the radiation power. However, in our
case, the nanoantenna has subwavelength dimensions and therefore it
can be neither classified as a hyperlens nor as a micro-spherical
dielectric nanoscope~\cite{Kim09,Lukyanchuk11}, moreover it is not
an analogue of solid immersion micro-lenses
~\cite{Rigneault08,Quake10,Hanson11,Wrachtrup10}, which are
characterized by the size 1-5 $\mu m$ in the same frequency range.
These lens has a subwavelength resolving power due to the large
geometric aperture but the value of normalized effective aperture is
$\mbox{S}_{n}\simeq1$. Our study demonstrates that the
sub-wavelength system, with \emph{small compared to the wavelength}
geometric aperture can have both high directing and resolving power
\emph{because of a strong increase of the effective aperture
compared to the geometrical one}.

\subsection{Experimental verification of superdirective optical
nanoantenna}

\begin{figure}[!t]
\centerline{\includegraphics[width=.8\textwidth]{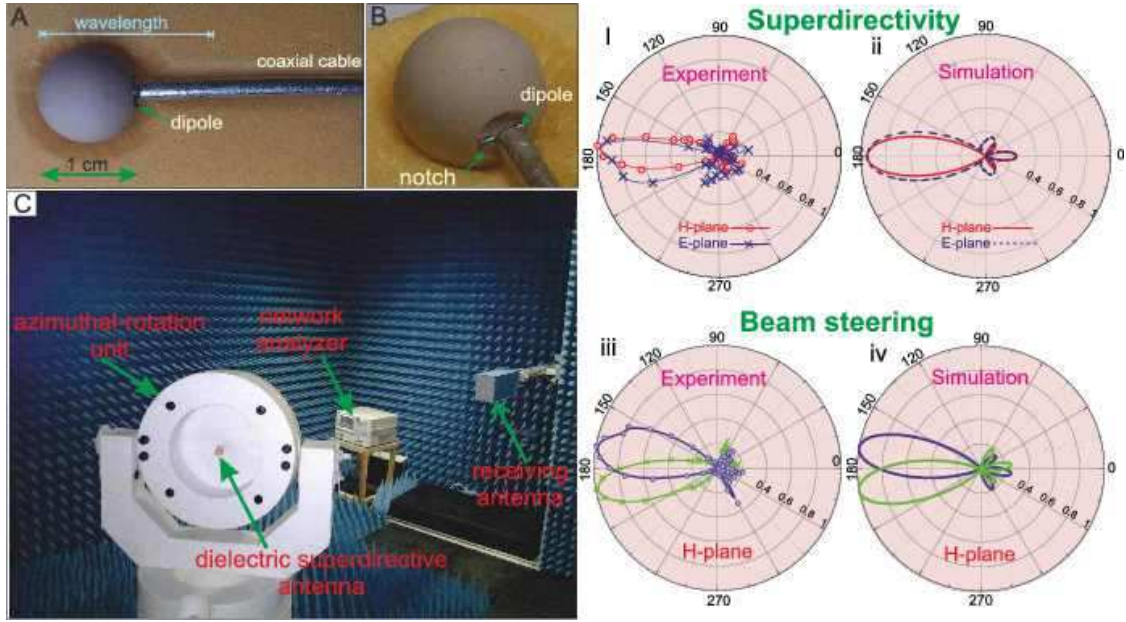}}
\caption{Photographs of ({\bf A}) top view and ({\bf B}) perspective
view of a notched all-dielectric microwave antenna. Image of ({\bf
C}) the experimental setup for measuring of power patterns.
Experimental (i) and numerical (ii) radiation patterns of the
antenna in both $E$- and $H$-planes at the frequency 16.8 GHz. The
crosses and circles correspond to the experimental data.
Experimental (iii) and numerical (iv) demonstration of beam steering
effect, displacement of dipole is equal 0.5 mm.} \label{Geomexper}
\end{figure}

We have confirmed both predicted effects studying the similar
problem for the microwave range. Therefore, we have scaled up the
nanoantenna as above to low frequencies. Instead of Si we employ
MgO-TiO$_{2}$ ceramic~\cite{8} characterized at microwaves by a
dispersion-less dielectric constant 16 and dielectric loss factor of
1.12$\cdot$10$^{-4}$. We have used the sphere of radius
$\mbox{R}_{\mbox{s}}=5$~mm and applied a small wire dipole~\cite{29}
excited by a coaxial cable as shown in Fig.~\ref{Geomexper}A,B.
The size of the hemispherical notch is approximately equal to
$\mbox{R}_{\mbox{n}}=2$ mm. Antenna properties have been studied in
an anechoic chamber Fig.~\ref{Geomexper}C.

The results of the experimental investigations and numerical
simulations of the pattern in both $E$- and $H$-planes are
summarized in Figs.~\ref{Geomexper}i,ii. Radiation patterns in
both planes are narrow beams with a lobe angle about 35$^{\circ}$.
Experimentally obtained coefficients of the directivity in both $E$-
and $H$-planes are equal to 5.9 and 8.4, respectively (theoretical
predictions for them were respectively equal 6.8 and 8.1). Our
experimental data are in a good agreement with the numerical results
except a small difference for the E plane, that can be explained by
the imperfect symmetry of the emitter. Note, that the observed
directivity is close to that of an all-dielectric Yagi-Uda antenna
with maximum size of $2\lambda$~\cite{8}. The maximum size of our
experimental antenna is closed to $\lambda/2.5$. Thus, our
experiment clearly demonstrates the superdirective effect.


Experimental and numerical demonstration of the beam steering effect
are presented in Figs.~\ref{Geomexper}iii,iv. For the chosen
geometry of antenna, displacement of source by 0.5 mm leads to a
beam rotation of about 10$^{\circ}$. Note that the ratio of
$\lambda=18.7$ mm to value of the source displacement 0.5 mm is
equal to 37. Therefore the beam steering effect observed at
subwavelength source displacement.

Finally, we consider the question of dielectric superdirective
antenna matching with coaxial cable. Despite that length of the wire
dipole is close to $\lambda/10$, dielectric superdirective antenna is well matched with the coaxial cable in the operating frequency range.
The antenna matching is explained by the strong coupling
of the wire dipole with the excited modes of notched dielectric
particle and is not related to the dissipative losses in the
superdirectivity regime. For this reason, we have not used
additional matching devices (e.g. "balun").

Though the concept of the superdirectivity of high-refractive index
dielectric particles with notch has now only been proven in GHz
spectral range there is a hope that it can be transferred into the
visible and near-IR spectrum in the nearest future. Recently we have
experimentally demonstrated that it is possible to engineer resonant
modes of spherical nanoresonators using a combined approach of
laser-induced transfer to generate almost perfect spherical
nanoparticles and helium ion beam milling to structure their surface
with sub-5nm resolution~\cite{Kuznetsovsplitball}. This novel
approach can become a suitable candidate for realizing
all-dielectric superdirective nanoantennas.

\section*{Conclusion}
\addcontentsline{toc}{section}{\textit{Conclusion}}

We propose a new type of highly efficient Yagi-Uda nanoantenna and
introduced a novel concept of superdirective nanoantennas based on
silicon nanoparticles. In addition to the electric response, this
silicon nanoantennas exhibit very strong magnetic resonances at the
nanoscale. Both types of nanoantennas are studied analytically,
numerically and experimentally. For superdirective nanoantennas we
also predict the effect of the beam steering at the nanoscale
characterized by a subwavelength sensitivity of the beam radiation
direction to the source position.

The unique optical properties and low losses make dielectric
nanoparticles perfect candidates for a design of high-performance
nanoantennas, low-loss metamaterials, and other novel all-dielectric
nanophotonic devices. The key to such novel functionalities of
high-index dielectric nanophotonic elements is the ability of
subwavelength dielectric nanoparticles to support simultaneously
both electric and magnetic resonances, which can be controlled
independently for particles of non spherical
forms.

\begin{spacing}{0.2}
\bibliographystyle{spiebib}
\addcontentsline{toc}{chapter}{References}

\end{spacing}

\end{sloppy}
\end{document}